\documentclass[11pt]{article}
\pdfoutput=1
\usepackage{jheppub}
\usepackage[T1]{fontenc}
\usepackage{cancel,tensor,enumitem,fix-cm}
\usepackage{epsfig}
\usepackage{graphicx}
\usepackage[english]{babel}
\usepackage{amsmath}
\usepackage{textcomp}
\usepackage{multirow}
\usepackage{booktabs}
\usepackage{tikz}
\usepackage{overpic}
\usepackage{bm}
\usepackage{physics}
\usepackage[lofdepth,lotdepth]{subfig}

\newcommand{\be}{\begin{equation}}
\newcommand{\ee}{\end{equation}}
\newcommand{\bea}{\begin{eqnarray}}
\newcommand{\eea}{\end{eqnarray}}
\def\({\left(}
\def\){\right)}
\def\[{\left[}
\def\]{\right]}

\def\p{\partial}
\def\si{\sigma}

\newsavebox{\imagebox}

\title{Towards bit threads in general gravitational spacetimes}

\author[a]{Dong-Hui Du}
\author[a]{and Jia-Rui Sun}
\affiliation[a]{School of Physics and Astronomy, Sun Yat-Sen University, Guangzhou, 510275, China}

\emailAdd{donghuiduchn@gmail.com}
\emailAdd{sunjiarui@mail.sysu.edu.cn}

\abstract{The concept of the generalized entanglement wedge was recently proposed by Bousso and Penington, which states that any bulk gravitational region $a$ possesses an associated generalized entanglement wedge $E(a) \supset a$ on a static Cauchy surface $M$ in general gravitational spacetimes, where $E(a)$ may contain an entanglement island $I(a)$. It suggests that the fine-grained entropy for bulk region $a$ is given by the generalized entropy $S_{\text{gen}}(E(a))$. Motivated by this proposal, we extend the quantum bit thread description to general gravitational spacetimes, no longer limited to the AdS spacetime. By utilizing the convex optimization techniques, a dual flow description for the generalized entropy $S_{\text{gen}}(E(a))$ of a bulk gravitational region $a$  is established on the static Cauchy surface $M$, such that $S_{\text{gen}}(E(a))$ is equal to the maximum flux of any flow that starts from the boundary $\partial M$ and ends at bulk region $a$, or equivalently, the maximum number of bit threads that connect the boundary $\partial M$ to the bulk region $a$. In addition, the nesting property of flows is also proved. Thus the basic properties of the entropy for bulk regions, i.e. the monotonicity, subadditivity, Araki-Lieb inequality and strong subadditivity, can be verified from flow perspectives by using properties of flows, such as the nesting property. Moreover, in max thread configurations, we find that there exists some lower bounds on the bulk entanglement entropy of matter fields in the region $E(a)\setminus a$, particularly on an entanglement island region $I(a) \subset (E(a)\setminus a)$, as required by the existence of a nontrivial generalized entanglement wedge. Our quantum bit thread formulation may provide a way to investigate more fine-grained entanglement structures in general spacetimes.}

\begin{document}
\maketitle
\flushbottom
%%%%%%%%%%%%%%%%%%%%%%%%%%%%%%%%%%%%%%%%%%%%%%%%%%%%%%%%%%%%%%%%%%%%%%%%%%%%%%
\section{Introduction}
%%%%%%%%%%%%%%%%%%%%%%%%%%%%%%%%%%%%%%%%%%%%%%%%%%%%%%%%%%%%%%%%%%%%%%%%%%%%%%
Relevant developments in the holographic nature of gravity \cite{tHooft:1993dmi,Susskind:1994vu} have revealed the deep connection between the spacetime geometry and the quantum entanglement. In the AdS/CFT correspondence \cite{Maldacena:1997re}, the Ryu-Tagayanagi (RT) formula \cite{Ryu:2006bv,Hubeny:2007xt} shows that the entanglement entropy of a boundary region $A$ is given by the area of a bulk minimal surface homologous to $A$, i.e.
\be \label{RT-formula}
S(A) = \min_{m\sim A} \frac{|m|}{4G_N} =\frac{|m_{A}|}{4G_N},
\ee
where $m$ is a bulk surface homologous to $A$ (represented by $m \sim A$), and $|m|$ represents taking the area of the surface $m$ for brevity. $m_{A}$ is the classical minimal surface homologous to $A$ in the bulk, which is termed as the RT surface. This formula is believed to hold at the order of $\mathcal{O}(1/G_N)$, in which the contribution from bulk quantum entanglement is ignored. When the leading-order quantum correction of order $\mathcal{O}(G_N^{0})$ coming from the bulk quantum entanglement is included, the RT formula should be corrected as the Faulkner-Lewkowycz-Maldacena (FLM) formula \cite{Faulkner:2013ana}, i.e.
\be \label{FLM-formula}
S(A) = \min_{m\sim A} \frac{|m|}{4G_N} + S_{\text{bulk}}(\si_A)  =  \frac{|m_A|}{4G_N} + S_{\text{bulk}}(\si_A)=S_{\text{gen}}(\si_A),
\ee
where $\si_A$ represents the spatial homology region surrounded by the union of surfaces $A\cup m_A$. $S_{\text{bulk}}(\si_A)$ is the von Neumann entropy of bulk matter fields restricted to $\si_A$. Note that in this formula, one needs first to take the minimization for the area term among all possible bulk surfaces $m \sim A$ to find the minimal surface $m_A$ and the associated homology region $\si_A$, then to add the bulk entanglement entropy of region $\si_A$. At the order of $\mathcal{O}(G_N^{0})$, the backreaction on the classical geometry due to the quantum corrections is ignored, so minimizing the area term exactly gives the classical RT surface. While it has been shown that the entanglement entropy with any higher order quantum corrections should be given by the known quantum extremal surface (QES) formula \cite{Engelhardt:2014gca}, that is
\be \label{QES-formula}
S(A) = \min_{m \sim A} S_{\text{gen}}(\si )= \min_{m \sim A} \[ \frac{|m|}{4G_N} + S_{\text{bulk}}(\si ) \]  =  \frac{|m_\mathcal{X}|}{4G_N} + S_{\text{bulk}}(\tilde{\si}_A) = S_{\text{gen}}(\tilde{\si}_A),
\ee
where $\si $ represents a spatial homology region with boundary $\p \si = A\cup m$ and $S_{\text{gen}}(\si )$ is the generalized entropy on region $\si $. Note that here $G_N$ is the renormalized Newton constant \cite{Susskind:1994sm} and $S_{\text{bulk}}$ is the finite part of the von Neumann entropy of bulk matter fields.\footnote{As discussed in \cite{Bousso:2015mna}, the UV divergence of the bulk entanglement entropy is precisely canceled by the renormalization of Newton's constant in the area term, and the sub-leading divergences are canceled by other geometric counter-terms. Therefore, the generalized entropy is argued to be finite and cutoff-independent.} Here we need to take the minimization for the generalized entropy (the union of the area term and the bulk entanglement term) among all possible bulk surfaces $m \sim A$ to find the so-called minimal quantum extremal surface $m_\mathcal{X}$, and $\tilde{\si}_A$ is a spatial homology region bounded by $A\cup m_\mathcal{X}$. The quantum extremal surface $m_\mathcal{X}$ is still a surface in the classical geometry, but it can significantly deviate from the classical  RT surface $m_{A}$. The so-called entanglement wedge \cite{Czech:2012bh,Wall:2012uf,Headrick:2014cta} $\text{EW}(A)$ is then defined as the bulk domain of dependence of the spatial region $\tilde{\si}_A$. It is possible for an entanglement wedge to contain a disconnected portion by definition, called the entanglement island. Recently, it has been shown that the QES formula plays a crucial role in solving the black hole information paradox. The island formula \cite{Penington:2019npb,Almheiri:2019psf,Almheiri:2019hni}, as a generalization from the QES formula, was proposed to calculate the entanglement entropy of the Hawking radiation and reproduce the Page curve \cite{Page:1993wv} in semi-classical gravity, which was verified by including the replica wormholes \cite{Penington:2019kki,Almheiri:2019qdq} in the gravitational path integral calculations in 2D Jackiw-Teitelboim (JT) gravity coupled to matters (for a nice review \cite{Almheiri:2020cfm}). It turns out that the entanglement wedge $\text{EW}(R)$ of a distant radiation region $R$ can contain an island region $I$ which includes most part of the black hole interior after the Page time, i.e. $\text{EW} (R)=R\cup I$. Then according to the entanglement wedge reconstruction \cite{Jafferis:2015del,Dong:2016eik,Cotler:2017erl}, it is possible for an observer who is restricted to region $R$ to recover the information on the region $I$ through sufficient complex operations. Subsequently, the entanglement island and the entanglement entropy have been widely studied in various gravitational spacetimes, such as AdS black hole spacetime coupled to non-gravitational bath \cite{Almheiri:2019yqk,Chen:2019uhq,Almheiri:2019psy,Tong:2023nvi,Lin:2024gip} and gravitational baths \cite{Geng:2020fxl,Anderson:2020vwi},
asymptotically flat black hole spacetime \cite{Hashimoto:2020cas,Hartman:2020swn,Alishahiha:2020qza,Matsuo:2020ypv,Wang:2021mqq,Yu:2021cgi,He:2021mst,Gan:2022jay,Du:2022vvg,Yu:2022xlh,Guo:2023gfa,Yu:2024fks}, de-Sitter spacetime \cite{Balasubramanian:2020xqf,Sybesma:2020fxg,Geng:2021wcq,Azarnia:2021uch}, AdS spacetime dual to the BCFT \cite{Rozali:2019day,Chen:2020uac,Geng:2020qvw,Geng:2021mic,Hu:2022ymx,Suzuki:2022xwv,Geng:2024xpj} and so on. More importantly, these studies indicated that the gravitational entropy formula may be applied in more general spacetimes.

On the other hand, the RT formula for entanglement entropy can be equivalently described by the bit thread formulation proposed by Freedman and Headrick \cite{Freedman:2016zud}, which can help clarify some conceptual puzzles around the RT formula. In the bit thread formulation, the entanglement entropy of a boundary subregion is given by the maximum flux of any flow (i.e. a divergenceless norm-bounded vector field) out of this boundary subregion, or equivalently the maximum number of bit threads (i.e. a set of integral curves of the vector field) with Planck-thickness that emanate from this boundary subregion to its complement region. With the developments of entropy formulas, the bit threads have been further developed to contain quantum corrections by properly modifying the divergenceless condition \cite{Chen:2018ywy, Agon:2021tia,Rolph:2021hgz}. Moreover, the bit threads have been generalized to the Lorentzian setting \cite{Headrick:2017ucz}, covariant setting \cite{Headrick:2022nbe} and higher curvature gravity \cite{Harper:2018sdd}. The bit threads have also been widely studied in holography to connect with other quantum information-theoretic concepts, such as the holographic entanglement of purification \cite{Agon:2018lwq,Du:2019emy,Bao:2019wcf,Harper:2019lff,Du:2019vwh,Lin:2020yzf}, holographic complexity \cite{Pedraza:2021mkh,Pedraza:2021fgp,Caceres:2023ziv}, holographic partial entanglement entropy \cite{Kudler-Flam:2019oru,Lin:2021hqs,Camargo:2022mme,Lin:2022aqf,Lin:2023orb,Lin:2023rxc,Lin:2024dho} and multipartite entanglement \cite{Harper:2020wad,Harper:2021uuq,Harper:2022sky}, which reveals many quantum information-theoretic aspects on the entanglement structures in holography. See Refs. \cite{Cui:2018dyq,Hubeny:2018bri,Agon:2019qgh,Agon:2020mvu,Headrick:2020gyq,Lin:2022flo,Gursoy:2023tdx,Lin:2023hzs,Caggioli:2024uza} for more recent studies on bit threads.

However, previous studies on the RT/FLM/QES formulas for holographic entanglement entropy and their corresponding bit thread formulations were mainly focused on the asymptotically AdS spacetimes.\footnote{The classical bit thread description was applied in dS space~\cite{Shaghoulian:2022fop,Susskind:2021esx}, as the RT prescription can be reformulated in de Sitter space in the framework of static-patch holography~\cite{Susskind:2021esx,Susskind:2021dfc,Shaghoulian:2021cef}.} Therefore, it is very important to find a gravitational entropy formula and its corresponding bit thread description that are applicable in more general spacetimes. For the entropy formula, recently, the entanglement wedge prescription has been extended to general spacetimes by Bousso and Penington \cite{Bousso:2022hlz,Bousso:2023sya}, called the generalized entanglement wedge (GEW), motivated by developments on the gravitational entropy
formulas \cite{Ryu:2006bv,Hubeny:2007xt,Faulkner:2013ana,Engelhardt:2014gca,Penington:2019npb,Almheiri:2019psf,Almheiri:2019hni,Penington:2019kki,Almheiri:2019qdq,Almheiri:2020cfm} and the tensor network toy models in quantum gravity \cite{Swingle:2009bg,Pastawski:2015qua,Hayden:2016cfa,Bao:2018pvs}.\footnote{Also motivated by tensor network models, the surface/state duality was proposed in \cite{Miyaji:2015yva,Miyaji:2015fia}, as an early attempt towards extending the holography into more general gravitational spacetimes. It says that in general gravitational spacetime $M$, a codimension-two convex surface $\Sigma_A \subset M$ (while the region $a$ considered in GEW proposal is codimension-one) is dual to a certain quantum state $\rho(\Sigma_A)$, and its entanglement entropy is given by the area of the minimal (quantum) extremal surface $m_A$ that is homologous to $\Sigma_A$. Thus it allows us to define an entanglement wedge $\text{EW}( \rho(\Sigma_A))$, i.e. the bulk domain of dependence of the spatial region bounded by $\Sigma_A\cup m_A$, in more general gravitational spacetimes. Another related progress was the proposal of a new approach for bulk reconstruction called the surface growth approach, in which the growth of general bulk extremal surfaces was used to reconstruct the bulk geometry and matter fields \cite{Lin:2020ufd,Yu:2020zwk,Fang:2024mwp}.} It has been shown that any bulk gravitational region $a$ posses an associated generalized entanglement wedge $E(a)$ on a static Cauchy surface $M$ in general spacetimes, where $E(a)$ is defined as the wedge that has the smallest generalized entropy $S_{\text{gen}}(E(a))$ among all wedges $E\supset a$ on $M$. Moreover, $E(a)$ may contain a disconnect entanglement island, denoted as $I(a)$. This proposal suggests that the fine-grained (von Neumann) entropy of bulk region $a$ is equal to $S_{\text{gen}}(E(a))$ by an entropy formula (\ref{genelized-QES}) similar to the QES formula (\ref{QES-formula}). While the QES prescription in AdS/CFT can be regarded as a special case of GEW prescription. In the present paper, we find that a direct application of GEW prescription in certain Cauchy surfaces would imply the so-called principle of the holography of information \cite{Laddha:2020kvp,Chowdhury:2020hse,Raju:2020smc,Raju:2021lwh}, which reveals the holographic nature of the boundaries of certain Cauchy slices. This prompts us to generalize the quantum bit threads into general gravitational spacetimes. To achieve this, we should note that the bit thread formulation was established by the convex optimization \cite{Freedman:2016zud,Headrick:2017ucz}, which can be applied to more general manifolds in principle, not limited to the AdS spacetime. As we will show in this work, the GEW proposal makes it feasible to extend the quantum bit thread formulation to general gravitational spacetimes. We will explore the GEW prescription from the bit thread perspectives, as the bit threads may provide more quantum information-theoretic aspects on the entanglement structures in general spacetimes.

The paper is organized as follows. In Section \ref{Sec:Qantm-bit-threads}, we briefly review the existing bit thread formulations corresponding to the RT/FLM/QES formulas in the AdS/CFT correspondence. In Section \ref{Sec:GEW}, we introduce the GEW proposal on a time-reflection symmetric Cauchy surface in general gravitational spacetimes, and we find that the GEW prescription would imply the principle of the holography of information on certain Cauchy surfaces in Section \ref{Sec:GEW-HOI}. In Section
\ref{Sec: Generalized bit threads}, we propose the quantum bit thread formulation that is dual to the entropy formula from the GEW proposal. We prove the dual bit thread formulation through the convex optimization in Section \ref{Sec:Dual quantum bit-thread}. Then in Section \ref{Sec:Properties of flows and entropy}, we first prove the nesting property of flows in Section \ref{Sec:nesting}, and then we prove the basic properties of entropy for bulk gravitational regions by using properties of flows in Section \ref{Sec:properties of entropy}. In section \ref{Sec:constraints from GEW}, we give an intuitive description of GEW prescription in terms of quantum bit threads, and we find that there exists nontrivial lower bounds on the bulk entanglement entropy of matter fields on region $E(a)\setminus a$, especially for an entanglement island region $I(a)$. The conclusion and discussion are given in Section \ref{Sec:Discussion}.

%%%%%%%%%%%%%%%%%%%%%%%%%%%%%%%%%%%%%%%%%%%%%%%%%%%%%%%%%%%%%%%%%%%%%%%%%%%%%%
\section{Review of the bit threads}\label{Sec:Qantm-bit-threads}
%%%%%%%%%%%%%%%%%%%%%%%%%%%%%%%%%%%%%%%%%%%%%%%%%%%%%%%%%%%%%%%%%%%%%%%%%%%%%%
Consider a time-reflection symmetric Cauchy surface $M$ with a conformal boundary $\partial M$, where $A$ is a subregion on $\partial M$. Define a divergenceless and norm-bounded flow $v$ on manifold $M$. The flux passing through region $A$ is given by
\begin{equation}\label{flux}
\int_A v :=\int_A\sqrt h\ n_{\mu}v^{\mu},
\end{equation}
where $h$ is the determinant of the induced metric $h_{ij}$ on $A$ and $n_{\mu}$ is the unit normal vector with inward-pointing direction. Then the entanglement entropy of region $A$ is given by the maximum flux of any flow through the region $A$, that is \cite{Freedman:2016zud}
\be\label{maxflow}
S(A) = \max_{v }\int_A v ,
\ee
subject to the constraints
\bea \label{flow-constraint}
\quad |v | \leq \frac{1}{4G_{N}},\  \nabla_{\mu}v^{\mu}  = 0.
\eea
The formula (\ref{maxflow}) was proved to be equivalent to the RT formula (\ref{RT-formula}) by using convex optimization and strong duality \cite{Headrick:2017ucz}. As there exits a max flow configuration $\tilde{v}$ subject to constraints (\ref{flow-constraint}) such that $\tilde{v}^\mu = n^\mu /4G_N$ at the classical minimal surface $m_A$, where the unit normal vector is outward-pointing on surface $m_A$. By using Gauss's law with the divergencelessness condition, it leads to
\be
S(A) = \int_A \tilde{v}= \int_{m_A} \tilde{v}= \frac{|m_{A}|}{4G_N}.
\ee

The early idea of adding quantum corrections to the bit thread prescription was discussed in \cite{Freedman:2016zud,Chen:2018ywy} by allowing the sources and sinks in the bulk or equivalently allowing bit threads to jump across the classical minimal surface. Then the so-called quantum bit thread formulation dual to the QES prescription was formally proposed in \cite{Agon:2021tia,Rolph:2021hgz} by using convex optimization similar to that in \cite{Headrick:2017ucz}, in which the divergenceless condition was modified properly in order to capture the contribution from the bulk entanglement of matter fields. The dual quantum bit thread formulation to all orders in $G_{N}$ was given by \cite{Rolph:2021hgz}
\be \label{quantum-maxflow}
S(A) = \max_{v }\int_A v ,
\ee\label{quantumflow-constraint}
subject to the constraints
\be \label{quantum-maxflow-constraint}
|v | \leq \frac{1}{4G_N},\ \forall\ \sigma  \in \Omega_A:   -\int_{\sigma } \nabla_{\mu}v^{\mu} \leq S_{\text{bulk}}(\sigma ) ,
\ee
where $\Omega_A$ represents the set of all possible homology regions for boundary region $A$, i.e.
\be
\Omega_A := \{ \sigma  \subseteq M: \p \sigma = A\cup m  \}.
\ee
There exists a max flow configuration $\tilde{v}$ subject to constraints (\ref{quantum-maxflow-constraint}) on $M$:
\begin{enumerate}
\item Requiring $\tilde{v}^\mu = n^\mu /4G_N$ at the quantum extremal surface $m_\mathcal{X}$, where $n^\mu$ is the unit normal vector on $m_\mathcal{X}$.
\item Saturating the divergence bound for $m=m_{\mathcal{X}}$ (thus $\si=\tilde{\si}_A$), that is
$-\int_{\tilde{\si}_A}  \nabla_\mu \tilde{v}^\mu = S_{\text{bulk}}(\tilde{\si}_A)$.
\end{enumerate}
The boundary of $\tilde{\si}_A$ has the orientation as $\p \tilde{\si}_A=m_\mathcal{X}-A$, where the unit normal vector is inward-pointing on surface $A$ and outward-pointing on surface $m_\mathcal{X}$. By using the Gauss's law, we have
\be
S(A) =\int_A \tilde{v}= \int_{m_\mathcal{X}} \tilde{v} - \int_{\tilde{\si}_A} \nabla_\mu \tilde{v}^\mu=\frac{|m_\mathcal{X}|}{4G_N} + S_{\text{bulk}}(\tilde{\si}_A),
\ee
as expected in the QES formula (\ref{QES-formula}). One may note that the quantum extremal surface $m_\mathcal{X}$ is generally not a minimal surface on the manifold $M$, thus it seems to violate the norm bound condition as the classical bit thread description shows that the norm bound can only saturate at the minimal surface. However, in the quantum bit thread description, the threads are no longer confined to the manifold $M$, they can jump out of the manifold at some points and then re-enter the manifold at other points, so that it is allowed to saturate the norm bound at a non-minimal surface. The validity of the norm bound condition has been checked in \cite{Rolph:2021hgz} near the quantum extremal surface.

\begin{figure}[h]
\centering
 \includegraphics[width=4.0 in]{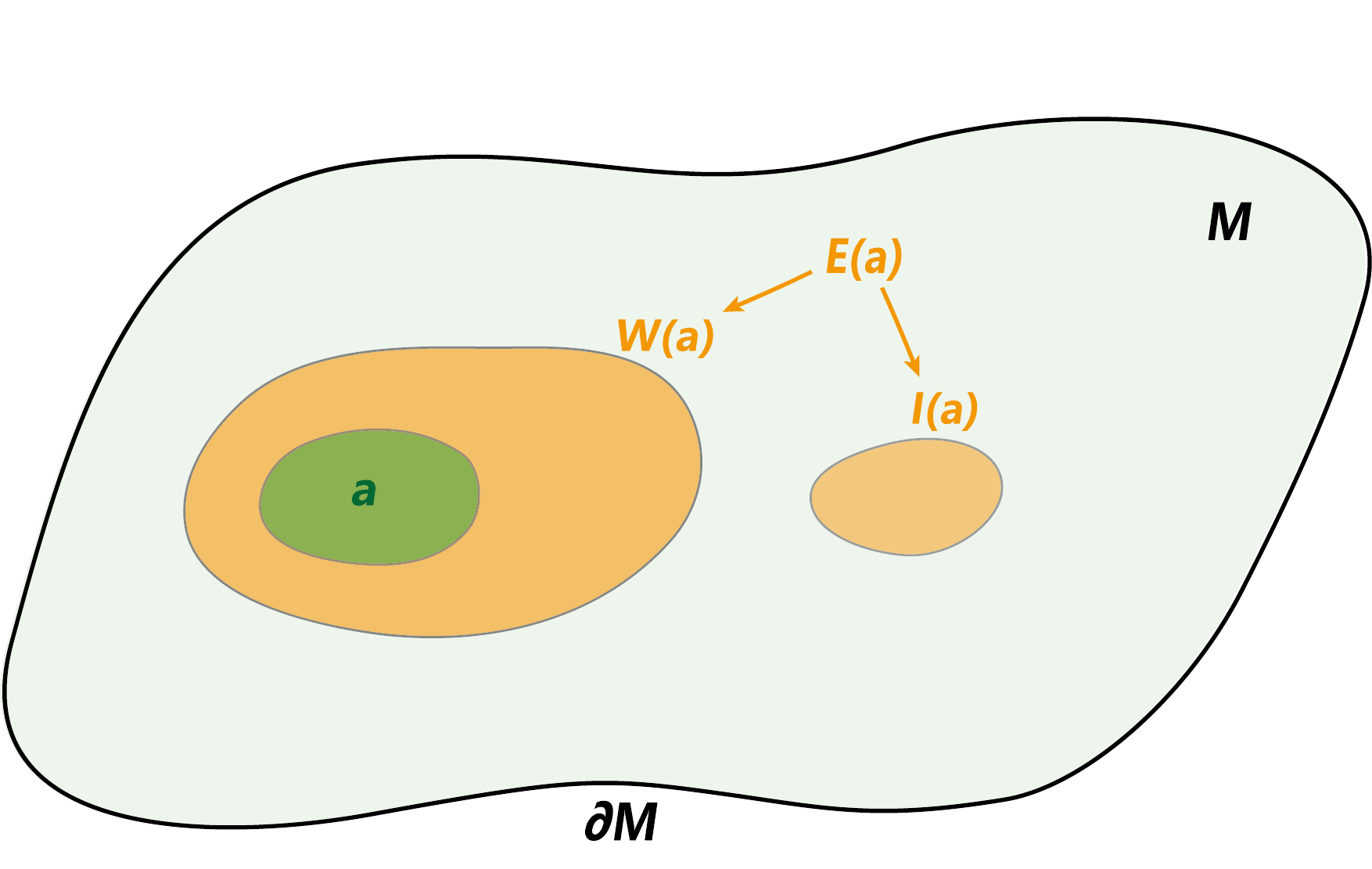}
     \caption{The GEW prescription on a static Cauchy surface $M$. Given any bulk region $a$ on $M$, there is an associated GEW $E(a)\supset a$, which minimizes the generalized entropy among all possible homology wedges that contain $a$. Moreover, $E(a)$ may contain a  island region $I(a)$, such that $E(a)=W(a)\cup I(a)$ with $W(a)\supset a$.}
     \label{Fig:GEW}
\end{figure}

%%%%%%%%%%%%%%%%%%%%%%%%%%%%%%%%%%%%%%%%%%%%%%%%%%%%%%%%%%%%%%%%%%%%%%%%%%%%%%
\section{GEW and generalized entropy for gravitational regions}\label{Sec:GEW}
%%%%%%%%%%%%%%%%%%%%%%%%%%%%%%%%%%%%%%%%%%%%%%%%%%%%%%%%%%%%%%%%%%%%%%%%%%%%%%
The entanglement wedge is crucial for understanding the holographic duality, which has been extended to general gravitational spacetimes by the GEW proposal \cite{Bousso:2022hlz,Bousso:2023sya}. For an arbitrary time-reflection symmetric Cauchy surface $M$ with asymptotic boundary $\p M$, i.e. a Riemannian manifold with metric $g_{\mu \nu}$. Denoting $\p s$ as the boundary of a set $s\subset M$, and defining $ \text{cl} \ s \equiv s \cup \p s$ and $\text{int}\ s = s \setminus \p s$, then \emph{wedge} $a$ is defined as any open subset of $M$ that is the interior of its closure: $ a = \text{int\ cl}\ a $. For two wedges $a$ and $b$, one can define the \emph{wedge union}, \emph{wedge complement} and \emph{wedge relative complement}, i.e.
\bea
    a \Cup b &\equiv & \text{int}\ \text{cl} (a \cup b)~,\\
    a' &\equiv & \text{int}\ a^{c}~, \\
    a \setminus b &\equiv & a \cap b'~,
\eea
which are also wedges, where $a^{c} \equiv M \setminus a$. The {\em generalized entropy} of any given wedge $a \subset M$ is defined as
\be \label{Sgen-definition}
    S_{\text{gen}}(a)\equiv \frac{ |\p a| }{4G_N}+ S_{\text{bulk}}(a),
\ee
where $G_N$ is the renormalized Newton constant, and $S_{\text{bulk}}(a) = - \tr \rho_a \log \rho_a$ is the finite part of the entanglement entropy of bulk matter fields, where $\rho_a = \tr_{a'} \rho$ is the density operator of matter fields on wedge $a$. The generalized entropy is finite and cutoff-independent \cite{Bousso:2015mna}. In this paper, we divide the total boundary of wedge $a$ into two parts as $\p a = \tilde{\p} a\cup \dot{a}$, where $\tilde{\p} a \equiv \p a \setminus \p M $ represents the part inside the bulk and $\dot{a} \equiv \p a \cap \p M$ represents the part that is overlapping with the boundary $\p M$.

According to the GEW proposal \cite{Bousso:2022hlz,Bousso:2023sya}, the GEW $E(a)$ associated to any given wedge $a$ is defined as the wedge that minimizes the generalized entropy among all possible wedges $\si  \supset a$ on $M$ (as illustrated in Figure \ref{Fig:GEW}). So the boundary $\p \si $ is defined on region $a^{c} = M \setminus a$ and it is homologous to $\p a$. Therefore, we have
\be \label{genelized-QES}
\begin{split}
S_{\text{gen}}(E(a))
&= \min_{\p \si \sim \p a \atop \text{on} \ a^{c}} S_{\text{gen}}(\si ) \\
&= \min_{\p \si  \sim \p a  \atop \text{on} \ a^{c}} \left[ \frac{|\p \si|}{4G_N} + S_{\text{bulk}}(\si ) \right]  \\
&= \frac{|\tilde{m}_\mathcal{X}|}{4G_N} + S_{\text{bulk}}(E(a)),
\end{split}
\ee
where we define $\tilde{m}_\mathcal{X}\equiv\p E(a)$ as a generalization of quantum extremal surface, distinguishing from $m_\mathcal{X}$ in the QES formula. It is also possible for the GEW to contain a disconnected portion, i.e. an entanglement island. This formula is suggested to be used to calculate the fine-grained (von Neumann) entropy of any codimension-one bulk region.  It reduces to the QES formula (more precisely, differing by a fixed area term $|A|$) when the bulk region $a$ is chosen to be a near-boundary region\footnote{More precisely, this near-boundary region $a$ can be defined as the union of the entanglement wedges of tiny boundary regions containing slightly smeared local boundary operators. As stressed in \cite{Bousso:2022hlz}, the local CFT operators in a conformal boundary region $A$ are dual to (quasi-) local bulk operators in the near-boundary bulk region $a$ \cite{Banks:1998dd,Hamilton:2006az}. So the algebra generated by bulk operators in the region $a$ is also able to encode the entire entanglement wedge of the region $A$, therefore $E(a)=\text{EW}(A)$ in this case. Notice that there exists an extra fixed area term $|A|$ in comparison to the QES formula, but it does not affect us to give the correct entanglement wedge prescription.} attached to the boundary subregion $A$ in AdS/CFT, where $E(a)=\text{EW}(A)$ and $\p E(a)= m_\mathcal{X}\cup A$. It should be emphasized here that in the formula (\ref{genelized-QES}) we need to take the minimization among all homologous surfaces $\p \si \sim \p a \sim \p M$, so it involves the entire boundary $\p M$ not just a subregion of the boundary, however, in QES formula we consider all homologous surfaces $m \sim A$, which only involves the subregion $A$ of the boundary $\p M$. Moreover, if we calculate the entanglement entropy of the Hawking radiation for evaporating black holes, we just set $a=R$, then we have $E(a)= R$ and $\p E(a)= \p R$ before the Page time, while $E(a)= R\cup I$ and $\p E(a)= \p R\cup \p I$ after the Page time, as given by the island formula.\footnote{As already mentioned in \cite{Hashimoto:2020cas}, the gravitational part of the generalized entropy should contain the area term of the boundary $\p R$ of the region $R$, which comes from the effect that region $R$ is separated from its complement. This also happens in the empty flat spacetimes. However, this term is often ignored, as it is fixed so it doesn't affect the minimization of the generalized entropy.} The GEW proposal suggests a generalization of entanglement wedge reconstruction such that the information in $E(a)$ can be reconstructed from region $a$.

%%%%%%%%%%%%%%%%%%%%%%%%%%%%%%%%%%%%%%%%%%%%%%%%%%%%%%%%%%%%%%%%%%%%%%%%%%%%%%
\subsection{Holography of information from GEW prescription}\label{Sec:GEW-HOI}
%%%%%%%%%%%%%%%%%%%%%%%%%%%%%%%%%%%%%%%%%%%%%%%%%%%%%%%%%%%%%%%%%%%%%%%%%%%%%%
Interestingly, we will point out that a direct application of GEW prescription in certain Cauchy surfaces would imply the principle of the holography of information \cite{Laddha:2020kvp,Chowdhury:2020hse,Raju:2020smc,Raju:2021lwh}, which claims that:
{\em In a theory of quantum gravity, a copy of all the information available on a Cauchy slice is also available near the boundary of the Cauchy slice. This redundancy in description is already visible in the low-energy theory}.

More specifically, we assume that $M$ is topologically trivial and the total quantum state of matter fields in Cauchy slice $M$ is a pure state. We choose the wedge $a$ to be a near-boundary bulk region $\delta_{M}$ attached to the asymptotic boundary $\p M$, with complement region $\delta_{M}^{c} \equiv M \setminus \delta_{M}$. Its boundary has the orientation $\p \delta_M= m_{\delta}-\p M$ (the unit normal vector is inward-pointing on $\p M$ and outward-pointing on $m_{\delta}$), where $m_{\delta}$ represents its inner boundary. According to the GEW proposal, we need to find the entanglement wedge $E(\delta_{M})$ that minimizes the generalized entropy, among all wedges $\si  \supset \delta_{M}$. The boundary of $\si$ is $\p \si=m - \p M $, hence $m \sim m_{\delta}\sim \p M$. As $\p M$ is fixed, we just need to vary the inner boundary $m$ on topologically trivial manifold $M$. Finally, one can find that the minimization leads to a trivial result, such that $m$ vanishes as a point and the entanglement wedge of the region $\delta_{M}$ is given by $E(\delta_{M})=M$. So the area contribution from the inner boundary $m$ vanishes, meanwhile $S_{\text{bulk}}(E(\delta_{M}))=S_{\text{bulk}}(M)=0$ as the total state of the matter fields in $M$ is assumed to be a pure state. The only contribution to entropy comes from the area term of the boundary $\p M$, which is a fixed term. That is
\be \label{holo-entropy}
\begin{split}
S_{\text{gen}}(E(\delta_{M}))
&= \min_{\p \si  \sim \p \delta_M  \atop \text{on} \ \delta_{M}^{c}} \[ \frac{|\p \si|}{4G_N} + S_{\text{bulk}}(\si ) \] \\
&= \frac{|\p M|}{4G_N}+ \min_{m  \sim m_{\delta}  \atop \text{on} \ \delta_{M}^{c}} \[ \frac{|\p m|}{4G_N} + S_{\text{bulk}}(\si ) \]= \frac{|\p M|}{4G_N}=S_{\text{gen}}(M).
\end{split}
\ee
As the entanglement wedge $E(\delta_{M})$ of the near-boundary bulk region $\delta_{M}$ contains the whole Cauchy surface $M$, the entanglement wedge reconstruction implies that all the information available on the Cauchy slice $M$ is also available near the boundary of the Cauchy slice, which leads to the principle of the holography of information. This reveals the holographic nature of boundaries of Cauchy slices, like the conformal boundary of AdS gravity.

In general, $M$ may have a nontrivial topology, and the quantum state of matter fields in $M$ may be mixed. In these general situations,
if we still choose $a$ to be the near-boundary bulk region $\delta_{M}$ that is attached to the outer asymptotic boundary (denoted as $\p M_{o}$ here), the entanglement wedge $E(\delta_{M})$ may have a nonvanishing interior boundary, meanwhile the entanglement entropy of matter fields on $E(\delta_{M})$ may not vanish. Hence $E(\delta_{M}) \neq M$ in general, there are some physical degrees of freedom outside $E(\delta_{M})$ that are not available for the observer restricted to the near-boundary region $\p M_{o}$. However, we stress that it is still possible to obtain a similar result like the formula (\ref{holo-entropy}) if we choose a proper near-boundary bulk region in general situations. To show this, there are two aspects that need to be noted. On the one hand, a topologically non-trivial $M$ can contain extra boundary $\p M_{i}$ (may consist of multiple separate inner boundaries). Hence the boundary of $M$ is given by $\p M=\p M_{i} \cup \p M_{o}$. We need to take $\p M_{i}$ into consideration as it may contain some nontrivial physical degrees of freedom. On the other hand, a mixed bulk quantum state in $M$ means that the matter fields in $M$ are entangled with matter fields in an extra manifold $M'$ (assuming that it is known and admits a semi-classical description) with boundary $\p M'$. However, as long as we consider the whole manifold $M_{\text{total}}=M\cup M'$ (whose topology may be complicated) with its complete boundary $\p M_{\text{total}}$, the total quantum state of matter fields in $M_{\text{total}}$ would still be pure. If we choose a near-boundary
bulk region $\delta_{M_{\text{total}}}$ that is attached to the whole boundary $\p M_{\text{total}}$, we expect $E(\delta_{M_{\text{total}}})= M_{\text{total}}$ to achieve the minimization of the generalized entropy. As the area term of the inner boundary of $E(\delta_{M_{\text{total}}})$ vanishes, meanwhile $S_{\text{bulk}}(E(\delta_{M_{\text{total}}}))=S_{\text{bulk}}(M_{\text{total}})=0$, with a remaining fixed area term $|\p M_{\text{total}}|$. This leads to a result similar to the formula (\ref{holo-entropy}).

\
%%%%%%%%%%%%%%%%%%%%%%%%%%%%%%%%%%%%%%%%%%%%
\section{Quantum bit threads in general gravitational spacetimes}\label{Sec: Generalized bit threads}
%%%%%%%%%%%%%%%%%%%%%%%%%%%%%%%%%%%%%%%%%%%%%%%%%%%%%%%%%%%%%%%%%%%%%%%%%%%%%%
In this section, we aim to find a quantum bit thread or flow description for the entanglement entropy of any spatial region in general gravitational spacetimes, which is dual to the entropy formula (\ref{genelized-QES}). Once we admit the holographic nature of the boundaries of certain Cauchy slices, we assume that the threads emerge from the boundaries of the Cauchy slice, as in the case of the AdS/CFT correspondence.

\begin{figure}[t!]
     \centering
     \includegraphics[width=0.65\textwidth]{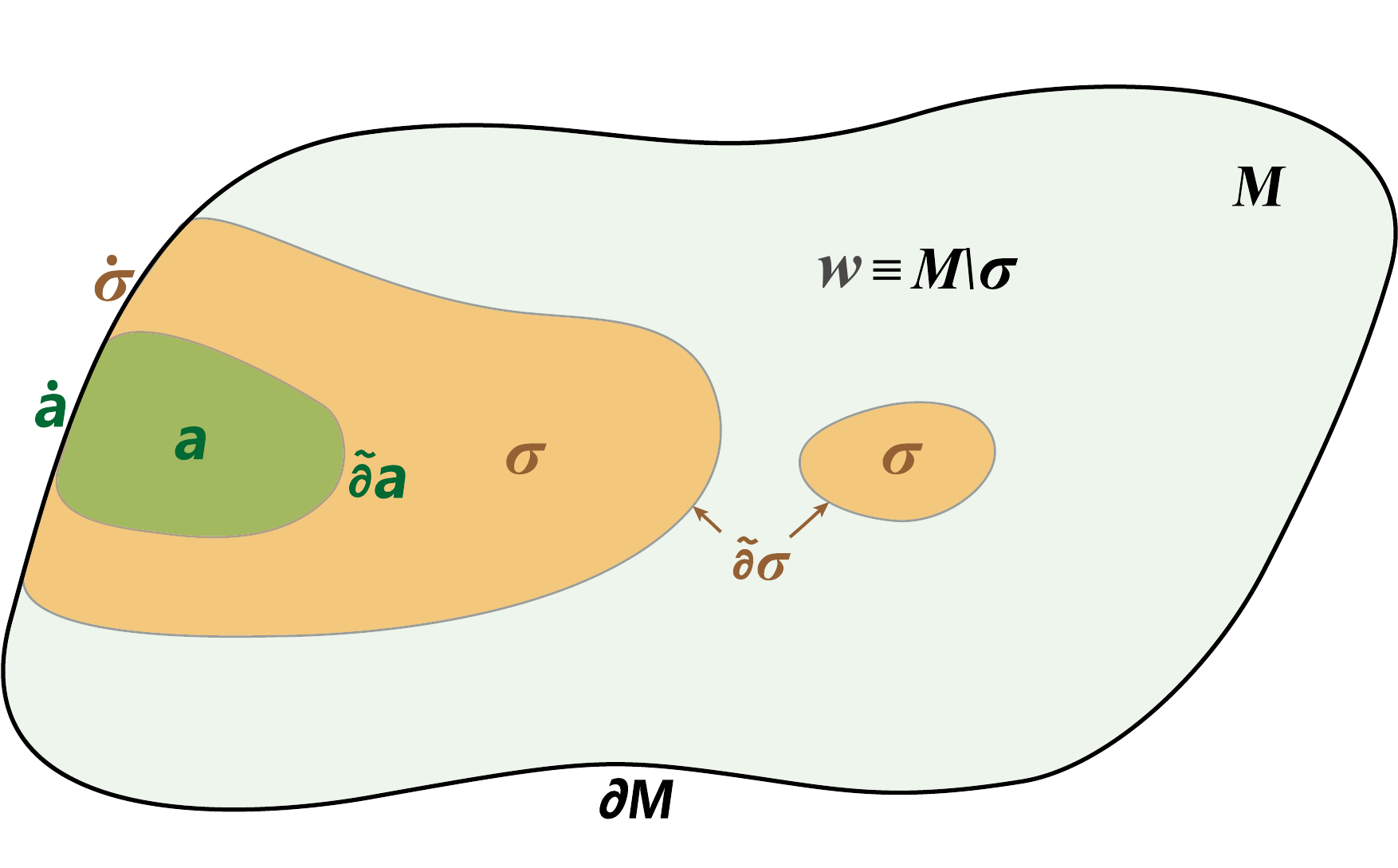}
     \caption{A schematic diagram for an allowed homology region $\si \supset a$ when $\dot{a} \neq \emptyset$. For $\dot{\si} \supset \dot{a}\neq \emptyset$, we have $m_{w} \equiv \p \si= \tilde{\p} \si \cup \dot{\si}=\tilde{\p} \si  \cup (\dot{\si} \setminus \dot{a}) \cup \dot{a}$, and we divide the entire boundary as $\p M=(\p M \setminus \dot{\si} )\cup \dot{\si}$. Therefore, $\p w= m_{w}- \p M = \tilde{\p} \si - (\p M\setminus \dot{\si})$ by canceling the overlapping part $\dot{\si}$. If $\dot{a} = \dot{\si}=\emptyset$, we will get $m_{w}\equiv \p \si = \tilde{\p} \si$ and $\p w=\tilde{\p} \si - \p M$. }
     \label{Fig:GEW-homology-region}
\end{figure}

In this paper, we consider an arbitrary time-reflection symmetric Cauchy surface $M$ (a topologically trivial manifold) with asymptotic boundary $\partial M$ (not necessary to be a conformal boundary), and we assume that the total state of matter fields in $M$ is a pure state. Given a wedge $a$ on the manifold $M$ with the boundary $\p a= \tilde{\p}a \cup \dot{a}$ (where $\dot{a}$ may be non-empty), so that $\tilde{\p}a \sim \dot{a}$. And $\si $ is a bulk homology region that contains wedge $a$, so that $a \subset \si \subset M$. If $\dot{a}\neq \emptyset$, by definition we must require $\dot{\si}\supset\dot{a}$, hence $(\p \si \setminus \dot{a})\sim \dot{a}$. Defining $w\equiv M\setminus \si$ as the complement of the region $\si$, due to the purity of the total state of matter fields on $M$, we have
\be \label{pure-relation}
S_{\text{bulk}}(\si) =S_{\text{bulk}}(w),
\ee
where the boundary of region $w$ is orientated as $\p w= m_{w}- \p M$, with $ m_{w} \equiv \p \si$ (which is allowed to partly overlap with $\p M$), as sketched in Figure \ref{Fig:GEW-homology-region}. So the formula (\ref{genelized-QES}) can be written as
\be \label{genelized-entropy-pure}
\begin{split}
S_{\text{gen}}(E(a))
&= \min_{\p \si  \sim \p a  \atop \text{on} \ a^{c}} \[ \frac{|\p \si|}{4G_N} + S_{\text{bulk}}(\si ) \]\\
&= \min_{ m_{w}  \sim \p M  \atop \text{on} \ a^{c}} \[ \frac{| m_{w}|}{4G_N} + S_{\text{bulk}}(w)  \] \\
&=\frac{| \tilde{m}_\mathcal{X}|}{4G_N} + S_{\text{bulk}}(\tilde{w}) ,
\end{split}
\ee
where the minimization is reached for $m_{w}=\p \si= \tilde{m}_\mathcal{X}$ when $w=\tilde{w}$ with boundary $ \p \tilde{w} = \tilde{m}_\mathcal{X}-\p M$. The generalized entanglement wedge $E(a)$ is just the bulk region $\tilde{\si}= M\setminus \tilde{w}$ that is surrounded by $\tilde{m}_\mathcal{X}$, with $S_{\text{bulk}}(\tilde{w})= S_{\text{bulk}}(\tilde{\si})$. For clarity, we see that the above minimization consists of two terms, i.e.
\bea \label{genelized-entropy-pure-1}
\frac{|\dot{a}|}{4G_N} + \min_{( m_{w} \setminus \dot{a} ) \sim (\p M \setminus \dot{a})  \atop \text{on} \ a^{c}} \[ \frac{|( m_{w} \setminus \dot{a} )|}{4G_N} + S_{\text{bulk}}(w) \].
\eea
If $ \dot{a} \neq \emptyset$, we have $(m_{w} \setminus \dot{a} ) \sim  \dot{a} \sim (\p M \setminus \dot{a})$, and then $S_{\text{bulk}}(w)$ reduces to the bulk entanglement entropy on the region that has the orientated boundary $(m_{w} \setminus \dot{a} ) - (\p M \setminus \dot{a})$. Note that this formula has the same form as the QES formula (\ref{QES-formula}) if we set $(m_{w} \setminus \dot{a})=m$ and $ (\p M \setminus \dot{a}) = A $ (differing by a fixed area term $|\dot{a}|=|A|$). For $\dot{a}=\emptyset$, it will be similar to the case with $A=\p M$ in the QES formula. Here a significant difference that should be stressed is that the minimization in formula (\ref{genelized-entropy-pure}) is only performed on region $a^{c}$, not the whole manifold $M$.

Therefore, based on the existing quantum bit thread descriptions, we expect that the entanglement entropy of a bulk region $a$ can be given by the maximum flux of any flow from the entire boundary $\p M$ to the bulk region $a$, or equivalently the maximum number of bit threads that start from the entire boundary $\p M$ and end at the bulk region $a$. The first area term in (\ref{genelized-entropy-pure-1}) is just dual to the part of the maximum flux directly entering the bulk region $a$ from the boundary $\p M$, as $\p a$ is just overlapping with $\p M$ at $\dot{a}$. While the second term in (\ref{genelized-entropy-pure-1}) is dual to the part of the maximum flux from $(\p M\setminus \dot{a})$ to the bulk region $a$, which can be similarly proved through the convex optimization techniques adopted for the QES formula with $(m_{w} \setminus \dot{a})=m$ and $ (\p M \setminus \dot{a}) = A $. It suggests that the entanglement entropy of any bulk region $a$ is equal to the maximum flux of any flow defined on $a^{c}=M\setminus a$ that starts from $\p M$ and then ends at region $a$. As we will prove in the next section (at a physicist's level of rigor), the quantum bit thread formulation dual to the entropy formula (\ref{genelized-entropy-pure}) can be formalized as
\be \label{quantum-maxflow-1}
S_{\text{gen}}(E(a)) = \max_{v_a  \atop \text{on} \ a^{c}}\int_{\p M} v_a = \max_{v_a  \atop \text{on} \ a^{c}} \(  \int_{\p a } v_a -\int_{ a^{c}} \nabla_\mu v^{\mu}_a \),
\ee
subjecting to the constraints
\be \label{flow-constraint-1}
|v_a | \leq \frac{1}{4G_N},\ \forall\ w  \in \Omega:   -\int_{w } \nabla_{\mu}v^{\mu}_a  \leq S_{\text{bulk}}(w ) ,
\ee
where $\Omega$ represents a set of the complements of all homology regions of region $a$, that is
\be \label{ homology region set}
\Omega := \{ w  \subset a^{c}: \p w =m_{w} - \p M \},
\ee
where the unit normal vector is inward-pointing on the boundary $\p M$ and outward-pointing on the surface $m_{w}$, and $m_{w} \sim \p a \sim \p M$. Note that the associated vector field $v_a$ is defined on region $a^{c}$, the dual bit thread program only needs the bulk metric and the bulk entanglement entropy of any bulk region $w$ on region $a^{c}$. There exists a max flow configuration $\tilde{v}_a$ which is subject to constraints (\ref{flow-constraint-1}) on $a^{c}$, satisfying:
\begin{enumerate}
\item $\tilde{v}^{\mu}_a = n^\mu /4G_N$ at the quantum extremal surface $\tilde{m}_\mathcal{X}$, where $n^{\mu}$ is the unit normal vector on $\tilde{m}_\mathcal{X}$.
\item For $m_{w}=\tilde{m}_\mathcal{X}$ (hence $w=\tilde{w}$), there is
$-\int_{\tilde{w}}  \nabla_\mu \tilde{v}^{\mu}_a = S_{\text{bulk}}(\tilde{w})$, where region $\tilde{w}$ has the orientated boundary $\p \tilde{w}=\tilde{m}_\mathcal{X}-\p M$.
\end{enumerate}
Therefore we have
\be  \label{flux of max-flow}
S_{\text{gen}}(E(a)) =\int_{\p M} \tilde{v}_a
=\int_{\p a } \tilde{v}_a -\int_{ a^{c}} \nabla_\mu
\tilde{v}^{\mu}_a
=\int_{\tilde{m}_{\mathcal{X}}} \tilde{v}_a -\int_{ \tilde{w}} \nabla_\mu
\tilde{v}^{\mu}_a
=\frac{|\tilde{m}_\mathcal{X}|}{4G_N} + S_{\text{bulk}}(\tilde{w}),
\ee
where Gauss's law is used. The dual flow description shows that a max flow achieves both the norm and divergence bounds for the quantum extremal surface $\tilde{m}_\mathcal{X}$ and its associated region $\tilde{w}$, which leads to the entropy formula (\ref{genelized-entropy-pure}). Moreover, we may divide any flow $v_{a}$ into the homogeneous part $v^{h}_{a}$ and inhomogeneous part $v^{i}_{a}$, such that
\be
\nabla v^{h}_{a}=0,\ \text{and}\ \forall\ w  \in \Omega: -\int_{w}  \nabla v^{i}_{a} \leq S_{\text{bulk}}(w)
\ee
on region $a^{c}$, and both the homogeneous and inhomogeneous parts can contribute to entropy.

\
%%%%%%%%%%%%%%%%%%%%%%%%%%%%%%%%%%%%%%%%%%%%%%%%%%%%%%%%%%%%%%%%%%%%%%%%%%%%%%
\subsection{Dual quantum bit thread formulation through the convex optimization}\label{Sec:Dual quantum bit-thread}
%%%%%%%%%%%%%%%%%%%%%%%%%%%%%%%%%%%%%%%%%%%%%%%%%%%%%%%%%%%%%%%%%%%%%%%%%%%%%%
In this section, we use the tools from the convex optimization, i.e. the Lagrangian duality and the strong duality, to prove the equivalence between the entropy formula (\ref{genelized-entropy-pure}) and the quantum bit thread formulation (\ref{quantum-maxflow-1}), similar to Refs. \cite{Headrick:2017ucz,Agon:2021tia,Rolph:2021hgz}. The Lagrangian duality is utilized to deal with a constrained optimization problem, which involves introducing Lagrange multipliers to enforce the constraints for a primal program, solving for the original variables, and obtaining an optimization problem for the Lagrange multipliers. Finally, the resulting dual program is equivalent to the primal program under certain conditions, as a result of strong duality.

A simple condition that implies strong duality is the Slater's condition \cite{Book:Convex-optimization}, which is crucial to the procedure of Lagrange duality. It requires that the primal problem admits a feasible point in the interior of the domain such that all the inequality constraints are strictly satisfied for this feasible point. Note that the formula (\ref{quantum-maxflow-1}) subject to the constraints (\ref{flow-constraint-1}) defines a concave program, in which the constraints (\ref{flow-constraint-1}) are concave functions of variable $v_a$. The Slater's condition is satisfied, as norm bound condition is the only non-linear constraint that needs to be strictly satisfied and $v_a = 0$ is a feasible point that strictly satisfies all the inequality constraints.

To obtain the dual convex program for the max flow program, we set $v_{a}$ as the original variable and introduce the Lagrange multipliers into each constraint. The Lagrangian function can be organized as
\be \label{Lagrangian-function}
\begin{split}
L[v_a,  \phi, \mu]
&= \int_{\p M} v_a + \int_{ a^{c}} \phi(x) \left(\frac{1}{4G_N}-|v_a|\right) + \int_{\Omega} d\mu (w) \[\left(\int_{  a^{c}} \chi (w,x)\nabla_\mu v_{a}^\mu\right) + S_{\text{bulk}}(w)\]\\
&=\int_{\p M} v_{a} + \int_{a^{c}} \phi(x) \left(\frac{1}{4G_N}-|v_{a}|\right) +\int_{\Omega} d\mu (w)S_{\text{bulk}}(w)\\
&\qquad + \int_{ a^{c}} \nabla_{\mu} \[ \(\int_{\Omega} d\mu(w) \chi (w ,x) \) v_{a}^{\mu} \]- \int_{a^{c}}v_{a}^{\mu} \p_{\mu}\(\int_{\Omega} d\mu(w) \chi (w ,x) \) \\
&= \int_{\p M} \left( \int_{\Omega} d\mu (w)  \bar{\chi} (w,x) \right) n_{\mu}v_{a}^{\mu} +  \int_{\p a} \( 1- \int_{\Omega} d\mu (w)  \bar{\chi} (w,x)\) n_{\mu}v_{a}^{\mu}  \\
&\qquad + \int_{\Omega} d\mu(w) S_{\text{bulk}}(w) + \int_{a^{c}} \[\phi(x)\left(\frac{1}{4G_N}-|v_{a}|\right)  + v_{a}^\mu \p_\mu \left(  \int_{\Omega} d\mu(w) \bar{\chi} (w ,x)\right) \],
\end{split}
\ee
where $\phi$ is a non-negative scalar field on $a^{c}$, $\mu$ is a non-negative probability measure on $\Omega$, such that $\int_{\Omega} d\mu(w) = 1$. Besides, we have performed the integral by parts in the second equality and used the definitions of characteristic functions $\chi (w,x)$ and $\bar{\chi}(w,x)$:
\be
\chi (w ,x) :=
\begin{cases}
1, \text{ for }x\in w \\
0, \text{ for }x\in a^{c}\backslash w
\end{cases}
\text{and} \ \ \ \chi (w ,x)+\bar{\chi} (w ,x) = 1
\ee
Then we maximize the Lagrangian function (\ref{Lagrangian-function}) with respect to the original variable $v_{a}$ to dualize the max flow program. Note that we impose no restrictions on the variable $v$ here, in order to make sure the finiteness of the maximization with respect to $v_{a}$, it demands that
\bea \label{finite-maximization-constraints}
\psi (x)|_{\p a^{c}} =  \bar{\chi} (\p a^{c},x):=
\begin{cases}
0, \text{ for }x\in \p M \\
1, \text{ for }x\in \p a.
\end{cases} \text{and} \ \ \ \phi(x) \geq \left|\p_{\mu} \psi (x) \right|,
\eea
where we have defined
\be\label{psi-func}
\psi (x) := \int_{\Omega} d\mu (w)  \bar{\chi} (w,x) \in [0,1].
\ee
With the constraint (\ref{finite-maximization-constraints}), the maximization of the Lagrangian function (\ref{Lagrangian-function}) reduces to
\be
\begin{split}
\max_{v_{a}} L[v_{a}, \phi, \mu]
&= \frac{1}{4G_N}\int_{a^{c}} \phi(x) + \int_{\Omega} d\mu (w) S_{\text{bulk}}(w).
\end{split}
\ee
Thus the dual Lagrangian is given by
\be \label{dual-program}
\begin{split}
\min_{\mu, \phi} \max_{v_{a}} L[v_{a}, \phi, \mu]
&= \min_{\mu} \left[ \frac{1}{4G_N}\int_{a^{c}} \left | \p_\mu \psi (x) \right | + \int_{\Omega} d\mu (w) S_{\text{bulk}}(w) \right],
\end{split}
\ee
which is subject to the constraint (\ref{finite-maximization-constraints}), in which the minimization  with respect to $\phi$ is trivial as $ \phi(x)$ can reach its minimum value, i.e. $ \phi(x)=\left|\p_{\mu} \psi (x) \right|$.

Now we argue that the result (\ref{dual-program}) is equivalent to the entropy formula (\ref{genelized-entropy-pure}). Given the function $\psi (x)$ on $a^{c}$ defined in (\ref{psi-func}) with the boundary condition (\ref{finite-maximization-constraints}), hence $ \psi(x) = 1$ at surface $\p a$ and  $ \psi(x) = 0$ at surface $\p M$. And we assume $\psi (x)$ is differentiable\footnote{Note that at surface $\dot{a}\equiv \p a \cap \p M\neq \emptyset$, the function $\psi(x)$ is non-differentiable by its boundary condition. However, the objective function only involves the gradient, i.e. $\p_{\mu}\psi(x)$. At surface $\dot{a}$, the result of this non-differentiable function can be defined as the limit of its value on a differentiable function. This argument is similar to the case with non-differentiability at $\p A$, referring to Footnote 10 in Ref. \cite{Headrick:2017ucz}. Moreover, it is feasible to use differentiable functions to approximate the non-differentiable function $\chi (w,x)$ well if we slightly smooth out the step function. Hence we can restrict to differentiable functions in our procedures.} on $a^{c}$. Then define a one-parameter family of bulk regions as
\be \label{one-parameter-rp}
r(p) := \{ x \in a^{c} : \psi(x) \leq p,\ \ p \in [0,1] \},
\ee
whose boundary is orientated as $\p r(p)=m(p)-\p M$, where $m(p)$ may partly overlap with the boundary $\p a^{c}=\p a-\p M$. By the continuity, we have $ \psi(x) = p$ at surface $ m(p)$. And we have $\p a \sim m(p) \sim \p M$ for $0\leq p\leq 1$, and $m(p)= \emptyset$ for $p<0$ or $p>1$.\footnote{The one-parameter family of bulk regions $r(p)$ corresponds to the set of the complements of all homology regions of $a$, i.e. set $\Omega$. One may define $ r'(p):= \{ x \in a^{c}: \psi(x) \geq p,\ p \in [0,1] \} $ as one-parameter family of bulk regions on $a^{c}$, whose boundary is $\p r'(p)= \p a- m(p)$. So that $a\cup r'(p)$ corresponds to the homology region of region $a$ on manifold $M$.}

As the normal derivative of a characteristic function is a surface delta function, thus the first term in the objective function (\ref{dual-program}) can be written in terms of the level sets as
\be \label{dual-area-term}
\begin{split}
\frac{1}{4G_N}\int_{a^{c}}  |\p_\mu \psi (x) |
= \frac{1}{4G_N}\int_{-\infty}^{\infty} dp\ |m(p)|= \frac{1}{4G_N}\int_0^1 dp\ |m(p)| ,
\end{split}
\ee
where $|m(p)|$ represents the area of $m(p)$. For the second term in the objective function (\ref{dual-program}), it can be argued to satisfy
\be \label{dual-bulk-term}
\int_{\Omega} d\mu (w) S_{\text{bulk}} (w) \geq \int_{0}^1 dp \ S_{\text{bulk}}(r(p)),
\ee
by taking advantage of the strong subadditivity of bulk entanglement entropy, i.e.
\be \label{bulk strong subadditivity} \sum_{i=1}^N S(a_i) \geq S(\cup_i a_i) + S(\cup_{\{i j\}} a_i \cap a_j ) + ... + S(\cap_i a_i),
\ee
as Ref. \cite{Rolph:2021hgz} did in its Argument 2. Here we choose the set of bulk regions $a_{i}$ as an arbitrary set of $N$ bulk homology regions $w_{i}$ with boundary $\p w_{i}= m_{i}-\p M$. The $n$th term on the right-hand side of inequality (\ref{bulk strong subadditivity}) is the bulk region $r(n/N)$, defined as $r(n/N) := \{ x \in a^{c} : \psi(x) \leq n/N \}$. Thus we have
\be
\sum_{i=1}^N S_{\text{bulk}} (w_i) \geq \sum_{i=1}^N S_{\text{bulk}}(r(i/N)),
\ee
which leads to formula (\ref{dual-bulk-term}) in the limit $N\to \infty$ after dividing by $N$. Finally, combining with
(\ref{dual-area-term}) and (\ref{dual-bulk-term}),
it gives the dual prescription
\be \label{dual-entropy-program}
\begin{split}
\min_{\mu, \phi}\max_{v_{a}} L[v_{a}, \phi, \mu]
=\min_{\mu} \int_0^1 dp \left( \frac{|m(p)|}{4G_N} + S_\text{bulk} (r(p))\right),
\end{split}
\ee
where $r(p)$ with its level set $m(p)$ is determined by measure $\mu$ on $\Omega$ defined in (\ref{ homology region set}). This is exactly the entropy formula (\ref{genelized-entropy-pure}). The minimization is realized if $\mu$ only supports on the homology region $\tilde{r}(p)$ that is bounded by the quantum extremal surface $\tilde{m}_{\mathcal{X}} $ and the boundary $\p M$, so that $\p \tilde{r}(p) = \tilde{m}_{\mathcal{X}} -\p M$. Meanwhile, the entanglement wedge time slice $E(a)$ of the region $a$ is defined as the bulk region that is surrounded by $\tilde{m}_{\mathcal{X}}$, thus $\p E(a)=\tilde{m}_{\mathcal{X}}$. Therefore, the proof on the equivalence between the entropy formula (\ref{genelized-entropy-pure}) and the bit thread formulation (\ref{quantum-maxflow-1}) subjecting to the constraints (\ref{flow-constraint-1}) is finished.

\
%%%%%%%%%%%%%%%%%%%%%%%%%%%%%%%%%%%%%%%%%%%%%%%%%%%%%%%%%%%%%%%%%%%%%%%%%%%%%%
\subsection{Nesting property and entropy properties for bulk gravitational regions  }\label{Sec:Properties of flows and entropy}
%%%%%%%%%%%%%%%%%%%%%%%%%%%%%%%%%%%%%%%%%%%%%%%%%%%%%%%%%%%%%%%%%%%%%%%%%%%%%%
With the quantum bit thread formulation for bulk gravitational regions, we will first prove the nesting property of flows. Then we are able to prove the basic properties of entropy (such as monotonicity, subadditivity, Araki-Lieb inequality, and strong subadditivity) for bulk gravitational regions from flow perspectives by using the nesting property of flows.

%%%%%%%%%%%%%%%%%%%%%%%%%%%%%%%%%%%%%%%%%%%%%%%%%%%%%%%%%%%%%%%%%%%%%%%%%%%%%%
\subsubsection{Nesting property of flows}\label{Sec:nesting}
%%%%%%%%%%%%%%%%%%%%%%%%%%%%%%%%%%%%%%%%%%%%%%%%%%%%%%%%%%%%%%%%%%%%%%%%%%%%%%
Let us prove the nesting property of flows on general Cauchy surface $M$. Given any two disjoint bulk regions $a$ and $b$ (i.e. $a\cap b = \emptyset$) on manifold $M$, we are interested in the total flux of any flow $v_{ab}$ defined on $M$ that starts from boundary $\p M$ and enters the union region $ab$ ($ab\equiv a\cup b$ for brevity). We assume that flow $v_{ab}$ consists of two independent components:
\be
v_{ab}=v_{a}+v_{b},
\ee
where $v_{a}$ is defined on region $a^{c}$ and it represents the flow component entering region $a$ (in both homogeneous and inhomogeneous ways), $v_{b}$ is defined on region $b^{c}$ and it represents the flow component entering region $b$ (in both homogeneous and inhomogeneous ways). We allow $v_{a}$ to pass through region $b$, but it can ``not end at'' region $b$,\footnote{Specifically, we mean that $(\int_{\p b} v_a -\int_{b} \nabla_\mu v^{\mu}_a )= 0$ by Gauss's law for flow $v_a$ on region $b$, where
flow $v_a$ is allowed to enter and leave region $a$.
Equivalently, it means that the corresponding oriented bit threads of $v_{a}$ (starting from the boundary $\p M$) are allowed to enter region $b$ in a homogeneous or inhomogeneous way, but it must leave from $b$ subsequently in a homogeneous or inhomogeneous way, as they must end at region $a$, not region $b$. This is what we mean ``not end at'' for flow $v_a$ on region $b$. It makes sure that flow $v_a$ has no contribution to entropy $S_{\text{gen}}(E(b))$ when we apply the quantum bit thread formulation for region $b$.} similarly for $v_{b}$. Thus $v_{a}$ contributes to $S_{\text{gen}}(E(a)) $ and $S_{\text{gen}}(E(ab)) $ but not to $S_{\text{gen}}(E(b)) $, meanwhile $v_{b}$ contributes to $S_{\text{gen}}(E(b)) $ and $S_{\text{gen}}(E(ab)) $ but not to $S_{\text{gen}}(E(a)) $. The nesting property of flows states that there exists a nesting max flow $v_{ab}=v_{a}+v_{b}$ that simultaneously maximizes the flux entering union region $ab$ and the flux entering region $a$ for nesting regions $a\subset ab$.

To prove the nesting property, first, we sum the flux that enters region $a$ and the flux entering union region $ab$, and we note that its maximum value should be bounded by the sum of several maximum values, i.e.
\bea \label{upper-bound-nesting}
\max_{v_{ab}} \left(\int_{\p M} v_{a} + \int_{\p M} v_{ab} \right )
\leq S_{\text{gen}}(E(a))  + S_{\text{gen}}(E(ab)) ,
\eea
which is subject to the norm bound constraints $|v_{a}|, |v_{ab}| \leq 1/4G_{N}$ and the divergence constraints for $v_a, v_{ab}$ on the complements of homology regions of region $a$ and region $ab$, respectively, denoted as $\Omega_1$ and $\Omega_2$, that is
\be \label{constraint-nesting}
\begin{split}
&\forall \  w \in \Omega_1: - \int_{w} \nabla_\mu v_{a}^\mu \leq S_{bulk}(w), \\
&\forall \  w \in (\Omega_1 \cup \Omega_2): - \int_{w} \nabla_\mu v_{ab}^\mu \leq S_{bulk}(w),
\end{split}
\ee
with $\Omega_{1} := \{ w  \subset a^{c}: \p w =m_{w} - \p M \}$ and $\Omega_{2} := \{ w  \subset (ab)^{c}: \p w =m_{w} - \p M \}$, where $\Omega_1 \supset \Omega_2$ due to the relation $a^{c}\supset (ab)^{c}$.

The nesting property can be proved as we will show that the value of the term on the left side of the inequality (\ref{upper-bound-nesting}) is also lower bounded by $S_{\text{gen}}(E(a))  + S_{\text{gen}}(E(ab)) $ by using the convex optimization. Let us construct the Lagrangian $L=L[v_{a},v_{b},\phi_{1},\phi_{2},\mu_{1},\mu_{2}]$ as follows
\be
\begin{split}
L =&  \int_{\p M} v_a + \int_{\p M} v_{ab} + \int_{a^{c}} \phi_{1} (x) \left ( \frac{1}{4G_N} - |v_{a}| \right ) + \int_{(ab)^{c}} \phi_{2} (x) \left ( \frac{1}{4G_N} - |v_{ab}| \right ) \\
& + \int_{\Omega_1} d\mu_{1}  \left(\int_{a^{c}} \chi (w, x) \nabla_\mu v_{a}^\mu+ S_{bulk}(w)\right)
+ \int_{\Omega_2} d\mu_{2}  \left(\int_{(ab)^{c}} \chi (w, x) \nabla_\mu v_{ab}^\mu+ S_{bulk}(w)\right)
\end{split}
\ee
in which $v_{a}, v_{b}$ as two independent original variables, $\phi_{1}$ and $\phi_{2}$ are two non-negative scalar fields, $\mu_{1}$ and $\mu_{2}$ are two probability measures on $\Omega_1$ and $\Omega_2$, respectively, satisfying $\int_{\Omega_1} d\mu_{1} (w)=\int_{\Omega_2} d\mu_{2} (w) = 1$. By performing the integral by parts, the Lagrangian can be organized as
\be
\begin{split}
L=& \int_{\p M} \left( \int_{\Omega_1} d\mu_{1}   \bar{\chi} (w,x) + \int_{\Omega_2} d\mu_{2}  \bar{\chi} (w,x)\right) n_{\mu}v_{a}^{\mu}
+ \int_{\p M} \(\int_{\Omega_2} d\mu_{2}  \bar{\chi} (w,x) \)n_{\mu}v_{b}^{\mu} \\
&+ \int_{\p a} \( 1- \int_{\Omega_1} d\mu_{1}   \bar{\chi} (w,x) - \int_{\Omega_2} d\mu_{2}   \bar{\chi} (w,x) \) n_{\mu}v_{a}^{\mu} + \int_{\p a} \( 1- \int_{\Omega_2} d\mu_{2}   \bar{\chi} (w,x)\)n_{\mu}v_{b}^{\mu} \\
&+ \int_{\p b} \(\int_{\Omega_2} d\mu_{2}   \bar{\chi} (w,x) \)n_{\mu}v_{ab}^{\mu} +  \int_{\Omega_1} d\mu_{1}  S_{\text{bulk}}(w)+\int_{\Omega_2} d\mu_{2} S_{\text{bulk}}(w) \\
&+ \int_{a^{c}} \[\phi_{1}(x)\left(\frac{1}{4G_N}-|v_{a}|\right)  + v_{a}^\mu \p_\mu \left(  \int_{\Omega_1} d\mu_{1} \bar{\chi} (w ,x)\right) \] \\
&+ \int_{(ab)^{c}} \[\phi_{2}(x)\left(\frac{1}{4G_N}-|v_{ab}|\right)  + v_{ab}^\mu \p_\mu \left( \int_{\Omega_2} d\mu_{2} \bar{\chi} (w ,x)\right) \]
\end{split}
\ee
Then we maximize the Lagrangian function with respect to the variables $v_{a},v_{b}$, while the finiteness of the maximization demands that
\bea
\psi_{1} (x)|_{\p a^{c}} =
\begin{cases}
0, \text{ for }x\in \p M \\
1, \text{ for }x\in \p a.
\end{cases} \text{and} \ \ \ \phi_{1}(x) \geq \left|\p_{\mu} \psi_{1} (x) \right|,\\
\psi_{2} (x)|_{\p (ab)^{c}} =
\begin{cases}
0, \text{ for }x\in \p M \\
1, \text{ for }x\in \p a.\\
1, \text{ for }x\in \p b.
\end{cases} \text{and} \ \ \ \phi_{2} (x) \geq \left|\p_{\mu} \psi_{2} (x)\right|,
\eea
where
\be\label{psi-function}
\psi_{1} (x) := \int_{\Omega_{1}} d\mu_{1} (w)  \bar{\chi} (w,x)\ \ \ \text{and} \ \ \ \psi_{2} (x) := \int_{\Omega_{2}} d\mu_{2} (w)  \bar{\chi} (w,x).
\ee
Therefore, we have
\be \label{dual-program-nesting}
\begin{split}
\min_{\phi_1,\phi_2,\mu_1,\mu_2} \max_{v_{a},v_{ab}} L
&= \min_{\mu_1,\mu_2} \left[ \frac{1}{4G_N} \( \int_{a^{c}} \left | \p_\mu \psi_{1} (x) \right | +  \int_{(ab)^{c}} \left | \p_\mu \psi_{2} (x) \right | \) + \int_{\Omega_1} d\mu_{1} S_{\text{bulk}}(w) \right. \\
&\left. \qquad\qquad\qquad\qquad \qquad\qquad\qquad\qquad\qquad\qquad\ \ \ + \int_{\Omega_2} d\mu_{2} S_{\text{bulk}}(w) \right]\\
&= \min_{\mu_1} \left[ \frac{1}{4G_N} \int_{a^{c}} \left | \p_\mu \psi_{1} (x) \right|  + \int_{\Omega_1} d\mu_{1} S_{\text{bulk}}(w) \right] \\
&\qquad\qquad\qquad\qquad \ \ + \min_{\mu_2} \left[ \frac{1}{4G_N} \int_{(ab)^{c}} \left | \p_\mu \psi_{2} (x) \right|  + \int_{\Omega_2} d\mu_{2} S_{\text{bulk}}(w) \right].
\end{split}
\ee
Further defining two independent sets of one-parameter family of bulk regions as
\bea \label{two one-parameter-rp}
& r(p_{1}):= \{ x \in a^{c} : \psi_{1} (x) \leq p_{1},\ \ p_{1} \in [0,1] \}, \nonumber \\
& r(p_{2}):= \{ x \in (ab)^{c} : \psi_{2} (x) \leq p_{2},\ \ p_{2} \in [0,1] \},
\eea
with the oriented boundaries $\p r(p_{1})=m(p_{1})-\p M,\ \p r(p_{2})=m(p_{2})-\p M$. According to Section \ref{Sec:Dual quantum bit-thread}, the objective function in formula (\ref{dual-program-nesting}) can be written in terms of two independent level sets, thus we obtain the following result
\be  \label{lower-bound-convex-duality}
\begin{split}
\min_{\phi_1,\phi_2,\mu_1,\mu_2} \max_{v_{a},v_{ab}} L
&\geq \min_{\mu_{1}} \int_0^1 d p_{1} \left( \frac{|m(p_{1})|}{4G_N} + S_\text{bulk} (r(p_{1}))\right)+\min_{\mu_{2}} \int_0^1 d p_{2} \left( \frac{|m(p_{2})|}{4G_N} + S_\text{bulk} (r(p_{2}))\right)  \\
&= S_{\text{gen}}(E(a)) +S_{\text{gen}}(E(ab)),
\end{split}
\ee
which leads to
\bea \label{lower-bound-nesting}
\max_{v_{ab}} \left(\int_{\p M} v_{a} + \int_{\p M} v_{ab} \right )
\geq S_{\text{gen}}(E(a))  + S_{\text{gen}}(E(ab)),
\eea
subject to the norm bound and divergence constraints around the formula (\ref{constraint-nesting}). Then combining with inequalities (\ref{upper-bound-nesting}) and (\ref{lower-bound-nesting}), which forces the inequalities to take the equal sign. It means that there exists an allowed max flow configuration $v_{ab}$, such that it simultaneously maximizes the flux entering union region $ab$ (with the maximal value $S_{\text{gen}}(E(ab))$) and the flux entering region $a$ (with the maximal value $S_{\text{gen}}(E(a))$), hence we finish the proof of the nesting property of flows.

\
%%%%%%%%%%%%%%%%%%%%%%%%%%%%%%%%%%%%%%%%%%%%%%%%%%%%%%%%%%%%%%%%%%%%%%%%%%%%%%
\subsubsection{Entropy properties for  bulk gravitational regions}\label{Sec:properties of entropy}
%%%%%%%%%%%%%%%%%%%%%%%%%%%%%%%%%%%%%%%%%%%%%%%%%%%%%%%%%%%%%%%%%%%%%%%%%%%%%%

\
\begin{itemize}
\item \textbf{Monotonicity}
\end{itemize}
The entropy for bulk regions $a, b$ satisfies the monotonicity property
\bea
S_{\text{gen}}(E(a))\leq S_{\text{gen}}(E(b)),\  \text{if}\  a\subset b.
\eea
Note that this property is not satisfied by the bulk entanglement entropy of matter fields. However, the GEW proposal shows that the union of the area term and the bulk entanglement entropy of matter fields should satisfy this property.

{\em Proof} : For bulk regions $a, b$, the dual flow description can be formalized as
\be\label{quantum-maxflow-ab}
\begin{split}
&S_{\text{gen}}(E(a)) = \max_{v \in \mathcal{F}}\int_{\p M} v, \
\mathcal{F} \equiv \{v\, \vert\  \forall\ w  \in \Omega: -\int_{w } \nabla_{\mu}v^{\mu} \leq S_{\text{bulk}}(w ),\ \ |v| \leq \frac{1}{4G_N} \}, \\
&S_{\text{gen}}(E(b)) = \max_{v \in \mathcal{F}'}\int_{\p M} v, \
\mathcal{F}' \equiv \{v\, \vert\  \forall\ w  \in \Omega': -\int_{w } \nabla_{\mu}v^{\mu} \leq S_{\text{bulk}}(w ),\ \ |v| \leq \frac{1}{4G_N} \},
\end{split}
\ee
with $\Omega := \{ w  \subset a^{c}: \p w =m_{w} - \p M \}$ and $\Omega' := \{ w  \subset b^{c}: \p w =m_{w} - \p M \}$. And the subscripts of $v_{a}, v_{b}$ are omitted for convenience. For $a\subset b$ with $b^{c}\subset a^{c}$, we have $\Omega' \subset \Omega$ and $\mathcal{F}'\subset \mathcal{F}$. Thus the set $\mathcal{F}$ has additional constraints than the set $\mathcal{F}'$. According to Theorem 1 in Section 2.4 of Ref. \cite{Harper:2019lff}, it leads to a constraint on solutions of two convex maximization programs in (\ref{quantum-maxflow-ab}), which gives the result as $S_{\text{gen}}(E(a))\leq S_{\text{gen}}(E(b))$ for $a\subset b$. Therefore, we always have $S_{\text{gen}}(E(a))\leq S_{\text{gen}}(E(ab))$ as $a\subset ab$ for arbitrary bulk regions $a, b$.

\begin{itemize}
\item \textbf{Subadditivity}
\end{itemize}
The entropy for any two disjoint bulk regions $a, b$ satisfies the subadditivity (SA) property
\bea
S_{\text{gen}}(E(ab))\leq S_{\text{gen}}(E(a))+ S_{\text{gen}}(E(b)).
\eea
{\em Proof}: We can begin with a nesting max flow $v_{ab}=v_a+v_b$ starting from boundary $\p M$ such that it simultaneously maximizes the flux entering the union region $ab$ and region $a$ for $a\subset ab$, as we proved before. It directly gives the SA property:
\bea
S_{\text{gen}}(E(ab))= \int_{\p M} v_{ab} = \int_{\p M} v_{a} + \int_{\p M} v_{b} \leq S_{\text{gen}}(E(a))+ S_{\text{gen}}(E(b)),
\eea
as the flux entering the union region $ab$ and region $a$ are equal to $S_{\text{gen}}(E(ab))$ and $ S_{\text{gen}}(E(a))$ respectively, while in general the flux entering the region $b$ can not reach its maximum value $S_{\text{gen}}(E(b))$.

\begin{itemize}
\item \textbf{Araki-Lieb inequality}
\end{itemize}
The entropy for bulk regions $a, b$ satisfies the Araki-Lieb (AL) property
\bea
|S_{\text{gen}}(E(a)) - S_{\text{gen}}(E(b))| \leq S_{\text{gen}}(E(ab))
\eea
{\em Proof} : A nesting max flow $v_{ab}$ that simultaneously maximizes the flux entering the union region $ab$ and region $a$, can also be used to prove this property. So that we have
\bea
S_{\text{gen}}(E(a))-S_{\text{gen}}(E(ab))=\int_{\p M} v_{a}- \int_{\p M} v_{ab} =  - \int_{\p M} v_{b} \leq S_{\text{gen}}(E(b)),
\eea
because the flux of any flow entering the region $b$ can not reach its maximum value $S_{\text{gen}}(E(b))$ in general, including the inverse flow $- v_{b}$. Similarly one can find $S_{\text{gen}}(E(b))-S_{\text{gen}}(E(ab)) \leq S_{\text{gen}}(E(a))$, thus the AL property is proved.

\begin{itemize}
\item \textbf{Strong subadditivity}
\end{itemize}
The entropy for any disjoint bulk regions $a, b, c$ satisfies the strong subadditivity (SSA) property
\bea
S_{\text{gen}}(E(b)) +S_{\text{gen}}(E(abc))\leq S_{\text{gen}}(E(ab))+S_{\text{gen}}(E(bc)).
\eea
{\em Proof} : We can start from a nesting max flow $v_{abc}=v_b+v_{ac}=v_a+v_b+v_c$ that simultaneously maximizes the flux entering the union region $abc$ and region $b$ as we have $b\subset abc$. It simply gives the SSA property:
\be
\begin{split}
S_{\text{gen}}(E(b)) +S_{\text{gen}}(E(abc))
&=\int_{\p M} v_{b}+ \int_{\p M} v_{abc}\\
&= \int_{\p M} v_{ab} +\int_{\p M} v_{bc} \\
&\leq S_{\text{gen}}(E(ab))+S_{\text{gen}}(E(bc))
\end{split}
\ee
as its flux entering the union region $abc$ and region $b$ are equal to $S_{\text{gen}}(E(abc))$ and $ S_{\text{gen}}(E(b))$ respectively, while in general the flux entering the union regions $ab$ and $bc$ can not reach their maximum values $S_{\text{gen}}(E(ab))$ and $S_{\text{gen}}(E(bc))$.

However, we would not expect that the monogamy of mutual information (MMI) property holds in general gravitational spacetimes. As the bulk gravity is not necessary to be AdS spacetime, the classical area term in the entropy formula (\ref{genelized-QES}) may not satisfy the MMI property like in AdS spacetime \cite{Hayden:2011ag,Headrick:2013zda}. Besides, the MMI property is not satisfied for general quantum states of bulk matter fields.\footnote{In paper \cite{Bousso:2024ysg}, it showed that the MMI property is satisfied for bulk gravitational regions if the bulk entanglement entropy can be neglected compared to the classical area term.}

%%%%%%%%%%%%%%%%%%%%%%%%%%%%%%%%%%%%%%%%%%%%%%%%%%%%%%%%%%%%%%%%%%%%%%%%%%%%%%
\section{Constraints from GEW prescription with bit threads}\label{Sec:constraints from GEW}
%%%%%%%%%%%%%%%%%%%%%%%%%%%%%%%%%%%%%%%%%%%%%%%%%%%%%%%%%%%%%%%%%%%%%%%%%%%%%%
In this section, we give an intuitive description of GEW prescription in terms of quantum bit threads. We will show that the existence of a nontrivial GEW will put constraints on the bulk entanglement entropy in certain bulk regions, such as regions with the entanglement island.

In addition to the flow description for the entropy formula from the GEW prescription given in Section \ref{Sec: Generalized bit threads}, it can also be described by a set of oriented and locally parallel discrete bit threads with Planck-thickness. The corresponding bit threads consist of the homogeneous part (with the number of threads denoted as $N^h$) as well as the inhomogeneous part (with the number of threads denoted as $N^i$). The homogeneous part represents the threads that are confined to the manifold $M$, like the description for original classical bit threads. These threads start from $\p M$, then pass through $\tilde{m}_\mathcal{X}$, and finally enter the region $a$ by passing through $\p a$. The number of these homogeneous threads is upper bounded by the area of the classical minimal surface that is homologous to $\p a$. While the inhomogeneous part represents the threads that can leave from the manifold $M$ at some points $P_{i}$ and re-enter the manifold $M$ at other points $Q_{i}$, whose distributions are restricted by bulk entanglement entropies for all regions $w\in \Omega$. One may think that the homogeneous threads are passing through ``classical cuts'' on manifold $M$, while inhomogeneous threads are passing through some ``quantum cuts''\footnote{The ``quantum cut'' just represents the part of cut (i.e. a codimension-2 surface) that is lying in extra geometries emerging from the quantum entanglement of bulk matter fields (for example, a cut on a micro wormhole emerging from an entangled EPR pair),  while the part of cut lying in original manifold $M$ is just called ``classical cut''. The emerging geometries are coupled to the original manifold $M$, thus forming an extended manifold with nontrivial topology. In the extended manifold, a cut that separates the whole manifold into two parts contains both the classical and quantum parts, as we call ``extended cut''. It suggests that an inhomogeneous thread passes through the ``quantum cuts'' in emerging geometries to complete a jump from some point to another point in original manifold $M$.} on extra geometries emerging from the bulk quantum entanglement \cite{VanRaamsdonk:2010pw}, such as $\text{ER}=\text{EPR}$  \cite{Maldacena:2013xja}. For each ``classical cut'' $m_{c}$, there is an associated ``quantum cut'' $m_{q}[m_{c}]$ that can be defined as a function of $m_{c}$, whose minimal area is equal to the bulk entanglement entropy between the matter fields on two sides of the surface $m_{c}$. In this way, the generalized entropy is described ``geometrically'' as an ``extended cut'' defined as $m_{\text{total}}[m_{c}]=m_{c}\cup m_{q}[m_{c}]$. Then by using a generalization of the Riemannian max-flow min-cut theorem \cite{Headrick:2017ucz}, the maximum number of total threads connecting the boundary $\p M$ to the bulk region $a$ through both ``classical cuts'' and ``quantum cuts'', is dual to the area of the minimal ``extended cut'' among all cuts $m_{\text{total}}[m_{c}]$.

\begin{figure}[h]
     \centering
\includegraphics[width=0.65\textwidth]{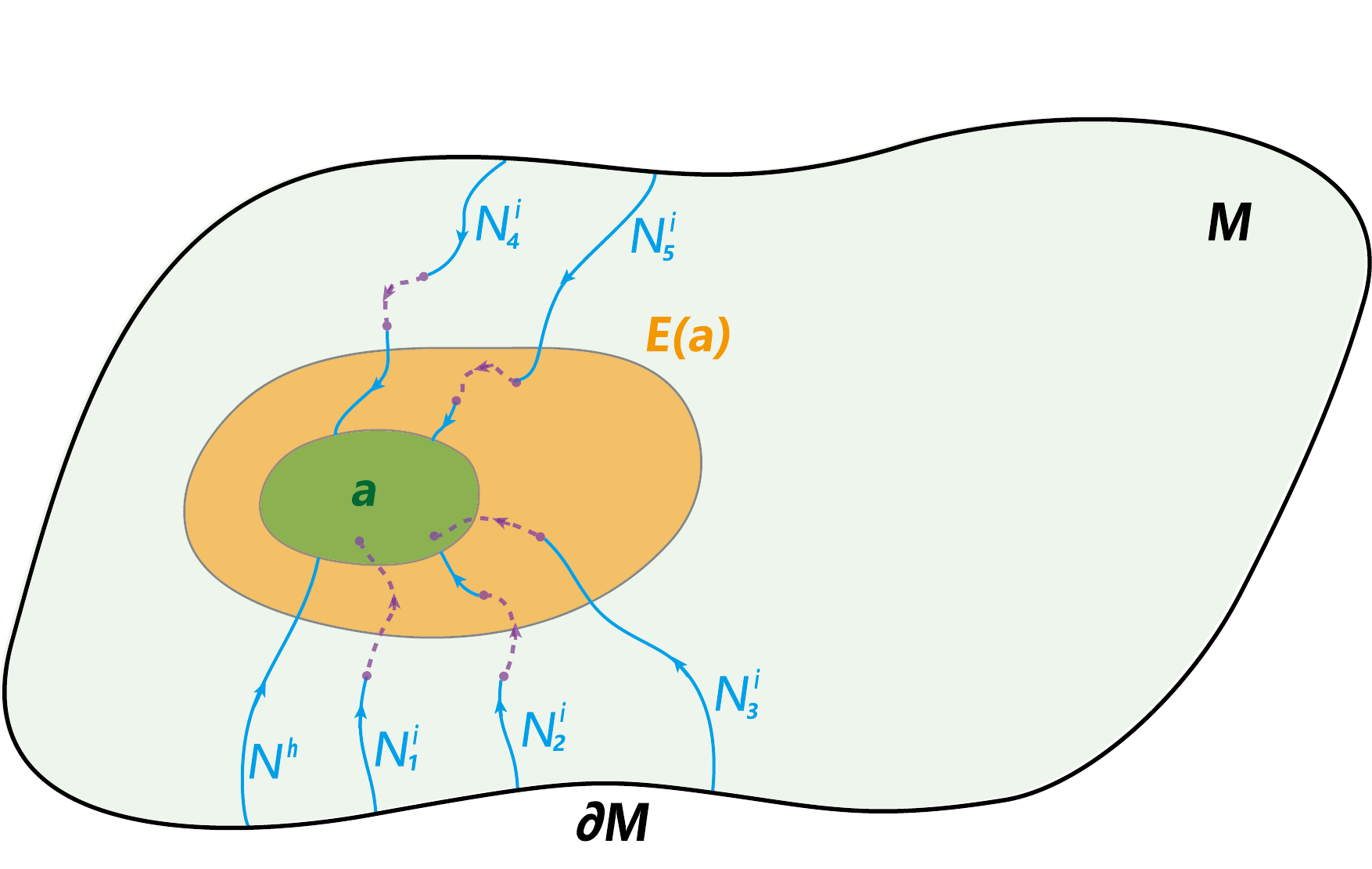}
     \caption{In a max thread configuration for GEW prescription without entanglement island.}
     \label{Fig:QBT-GEW-no island}
\end{figure}

According to the GEW proposal, for a given bulk region $a$ on Cauchy surface $M$, it is possible to make the total generalized entropy smaller by including extra regions besides $a$ as a part of $E(a)$, such that $E(a)=a \cup (E(a)\setminus a)$, where $E(a)\setminus a$ may be nonempty. Compared to region $a$, we are able to make the bulk quantum state of matter fields in $E(a)$ purer by including more bulk entangled partners into $E(a)$. So the bulk entanglement entropy term in generalized entropy can be decreased, but the risk is that the boundary area term may be increased. Thus, we need make sure that $S_{\text{gen}}(E(a))\leq S_{\text{gen}}(a)$ by definitions. As shown in Figure \ref{Fig:QBT-GEW-no island}, for the case that $E(a)$ contains no entanglement island, the bulk region $M$ is divided into three parts as $M= a \cup (E(a)\setminus a) \cup (M\setminus E(a))$.
Based on the ways of the inhomogeneous threads passing through the surface $\p a$ and $\p E(a)$, they can be further divided into five types. These oriented threads start from $\p M$, but they can directly jump into region $a$ from region $M\setminus E(a)$ (with the number of threads $N_{1}^i$), or jump into region $E(a)\setminus a$ from region $M\setminus E(a)$ and then enter region $a$ by passing through $\p a$ (with the number of threads $N_{2}^i$), or enter into region $E(a)\setminus a$ by passing through $\p E(a)$ and then jump into $a$ from region $E(a)\setminus a$ (with the number of threads $N_{3}^i$). And there are two more cases where $P_{i}$ and $Q_{i}$ are located in the same region $M\setminus E(a)$ (with the number of threads $N_{4}^i$) or region $E(a)\setminus a$ (with the number of threads $N_{5}^i$), these inhomogeneous
threads pass through both surfaces $\p a$ and $\p E(a)$. Now consider a max flow configuration for the no-island phase, satisfying
\begin{enumerate}
\item[(A1)] $\tilde{v}^{\mu}_a = n^\mu /4G_N$ at the quantum extremal surface $\tilde{m}_\mathcal{X}=\p E(a) $, with the unit normal vector $n^{\mu}$ on $\tilde{m}_\mathcal{X}$.
\item[(A2)] For $m_{w}=\tilde{m}_\mathcal{X}$ (hence $w=\tilde{w}=M\setminus E(a)$), there is
$-\int_{\tilde{w}}  \nabla_\mu \tilde{v}^{\mu}_a = S_{\text{bulk}}(\tilde{w})=S_{\text{bulk}}(E(a))$, where region $\tilde{w}$ has the orientated boundary $\p \tilde{w}=\tilde{m}_\mathcal{X}-\p M$
\end{enumerate}
In other words, it requires that in a max thread configuration, the part of oriented threads entering $E(a)$ by passing through $\p E(a)$ reaches its maximum number, i.e. $|\p E(a)|/(4G_{N})$. Meanwhile, the part of oriented threads entering $E(a)$ by jumping from $\tilde{w}$ to $E(a)$ also reaches its maximum number, i.e. $ S_{\text{bulk}}(E(a))$. While for other surfaces $m_{w}\neq \tilde{m}_\mathcal{X}$ and its associated bulk region $w$, these two parts of threads can not reach their maximum simultaneously. Therefore, in a max thread configuration, we have
\be
\begin{split}
S_{\text{gen}}(E(a))&=\frac{|\p E(a)|}{4G_N} + S_{\text{bulk}}(E(a))=(N^{h}+N^{i}_{3}+N^{i}_{4}+N^{i}_{5})+(N^{i}_{1}+N^{i}_{2}),\\
S_{\text{gen}}(a)&=\frac{|\p a|}{4G_N} + S_{\text{bulk}}(a)\geq (N^{h}+N^{i}_{2}+N^{i}_{4}+N^{i}_{5})+(N^{i}_{1}+N^{i}_{3}),
\end{split}
\ee
where the number of each type of threads is non-negative. Thus the inequality $S_{\text{gen}}(E(a))\leq S_{\text{gen}}(a)$ is satisfied automatically. It leads to the constraint
\be
\begin{split}
S_{\text{bulk}}(a)-S_{\text{bulk}}(E(a))\geq N^{i}_{3}-N^{i}_{2} \geq \frac{|\p E(a)|-|\p a|}{4G_N}.
\end{split}
\ee
Furthermore, we can find a lower bound on the bulk entanglement entropy of region $E(a)\setminus a$ with orientated boundary $\p (E(a)\setminus a)=\p E(a)- \p a$, that is
\be \label{constraint1}
\begin{split}
S_{\text{bulk}}(E(a)\setminus a)\geq S_{\text{bulk}}(a)-S_{\text{bulk}}(E(a))\geq N^{i}_{3}-N^{i}_{2} \geq \frac{|\p E(a)|-|\p a|}{4G_N},
\end{split}
\ee
where we have assumed the Araki-Lieb inequality of the bulk entanglement entropy for the first inequality, as $E(a)=a \cup (E(a)\setminus a) $. It puts a constraint on the existence of a nontrivial GEW for cases without island.

\begin{figure}[t!]
     \centering
\includegraphics[width=0.65\textwidth]{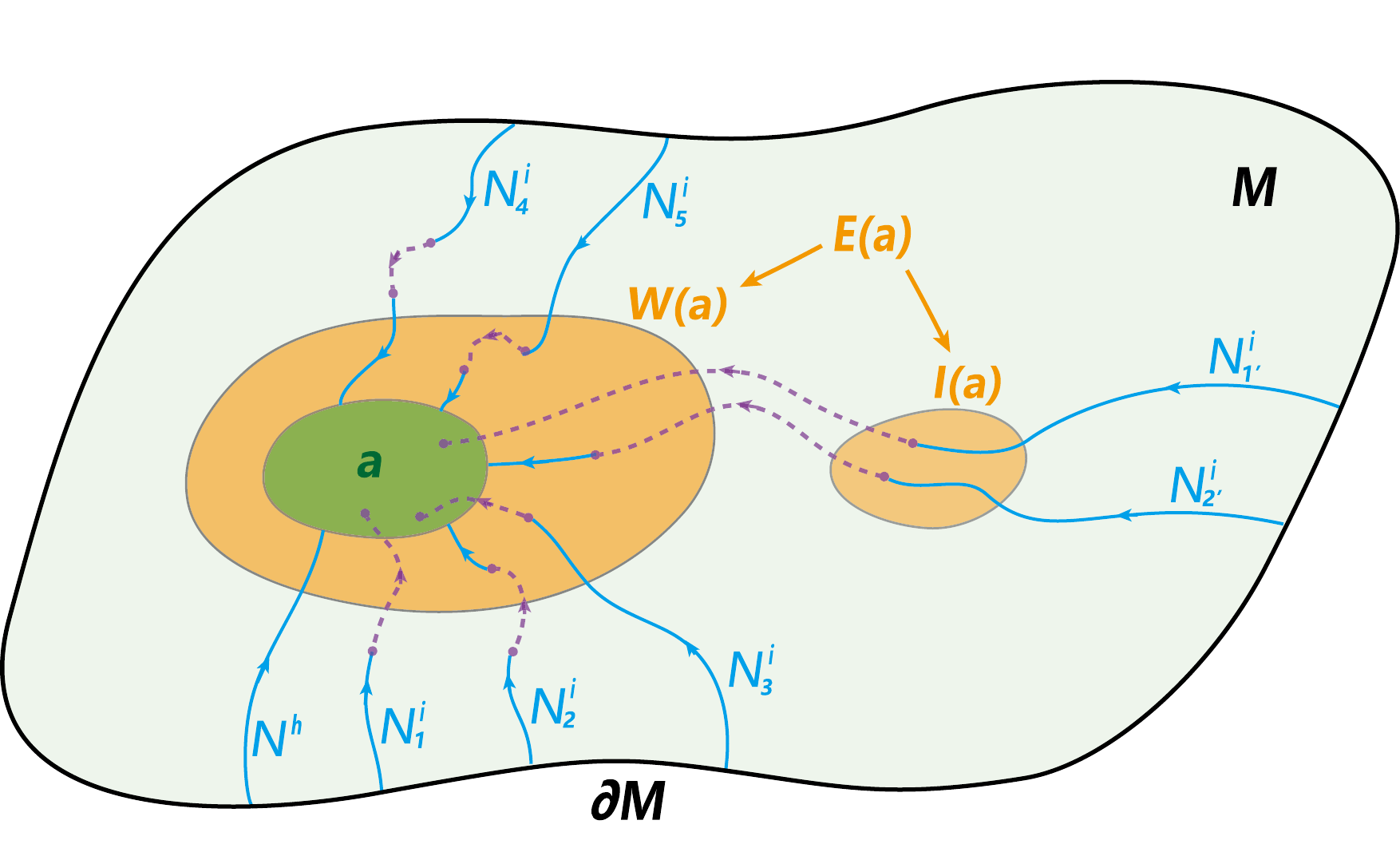}
     \caption{In a max thread configuration for GEW prescription with an entanglement island.}
     \label{Fig:QBT-GEW-with island}
\end{figure}

It is interesting to consider the case when $E(a)$ contains an entanglement island, such that $E(a)=W(a)\cup I(a)$, as shown in Figure \ref{Fig:QBT-GEW-with island}. By introducing an extra disconnected region $I(a)$ as a part of $E(a)$, it is possible to further purify the bulk quantum state of matter fields in $E(a)$, as we are able to include more bulk entangled partners into $E(a)$. This makes the bulk entanglement entropy term decrease but at the cost of increasing a boundary area term of the island. For the island phase, we need make sure that $S_{\text{gen}}(E(a))\leq \{ S_{\text{gen}}(W(a)),\ S_{\text{gen}}(a) \}$  by definitions. Recall that when there is no island region, there are six types of threads connecting boundary $\p M$ to region $a$, with numbers $N^h, N_{1}^i,N_{2}^i,N_{3}^i,N_{4}^i,N_{5}^i$. Then after introducing island $I(a)$, all these six types of threads may pass through $I(a)$. Now we consider a max flow configuration for the island phase, which satisfies
\begin{enumerate}
\item[(B1)] $\tilde{v}^{\mu}_a = n^\mu /4G_N$ at the quantum extremal surface $\tilde{m}_\mathcal{X}=\p E(a)=\p W(a) \cup \p I(a)$, with the unit normal vector $n^{\mu}$ on $\tilde{m}_\mathcal{X}$.
\item[(B2)] For $m_{w}=\tilde{m}_\mathcal{X}$ (hence  $w=\tilde{w}=M\setminus E(a)$), there is
$-\int_{\tilde{w}}  \nabla_\mu \tilde{v}^{\mu}_a = S_{\text{bulk}}(\tilde{w})=S_{\text{bulk}}(E(a))$, where region $\tilde{w}$ has the orientated boundary $\p \tilde{w}=\tilde{m}_\mathcal{X}-\p M$.
\end{enumerate}
Therefore, in a max thread configuration, the part of oriented threads entering $E(a)$ by passing through $\p E(a)$ reach its maximum number, i.e. $(|\p W(a)|+|\p I(a)|)/(4G_{N})$. And the part of oriented threads entering $E(a)$ by jumping from region $\tilde{w}$ to region $E(a)$ also reaches its maximum number, i.e. $ S_{\text{bulk}}(E(a))$. Now we focus on region $I(a)$, as the part of oriented threads entering $I(a)$ by passing through $\p I(a)$ reaches its maximum number, i.e. $|\p I(a)|/(4G_{N})$. It requires these oriented threads can only leave region $I(a)$ by jumping across $\p I(a)$ without passing through  $\p I(a)$, otherwise the part of flux entering $I(a)$ by passing through $\p I(a)$ will not reach its maximum if these oriented threads are allowed to turn back by passing through $\p I(a)$ again (with negative contributions to the flux passing through $\p I(a)$ with orientation). Moreover, for those oriented threads that leave region $I(a)$ by jumping across surface $\p I(a)$, they can only jump into region $W(a)$ instead of region $\tilde{w}$. As the second condition requires that the part of flux jumping from region $\tilde{w}$ to region $E(a)$ also reaches its maximum, thus it does not allow these oriented threads to jump inversely from region $I(a)$ to region $\tilde{w}$ (with negative contributions to the flux jumping from region $\tilde{w}$ to region $E(a)=W(a)\cup I(a)$ with orientation). Finally, it turns out that not all six types of threads can be assigned to the island, there are only two types of threads that are allowed to be assigned to $I(a)$ for a max thread configuration, with numbers denoted as $N_{1'}^i,N_{2'}^i$ in Figure \ref{Fig:QBT-GEW-with island}. Therefore, in a max thread configuration, we have
\be
\begin{split}
S_{\text{gen}}(E(a))&=\frac{|\p W(a)|}{4G_N} +\frac{|\p I(a)|}{4G_N}+ S_{\text{bulk}}(W(a)\cup I(a))\\
&=(N^{h}+N^{i}_{3}+N^{i}_{4}+N^{i}_{5})+(N^{i}_{1'}+N^{i}_{2'})+(N^{i}_{1}+N^{i}_{2}),\\
S_{\text{gen}}(W(a))&=\frac{|\p W(a)|}{4G_N} + S_{\text{bulk}}(W(a))\\
&\geq (N^{h}+N^{i}_{3}+N^{i}_{4}+N^{i}_{5})+(N^{i}_{1}+N^{i}_{2}+N^{i}_{1'}+N^{i}_{2'}),\\
S_{\text{gen}}(a)&=\frac{|\p a|}{4G_N} + S_{\text{bulk}}(a)\\
&\geq (N^{h}+N^{i}_{2}+N^{i}_{2'}+N^{i}_{4}+N^{i}_{5})+(N^{i}_{1}+N^{i}_{3}+N^{i}_{1'}),
\end{split}
\ee
where the number of each type of threads is non-negative. Thus the inequalities $S_{\text{gen}}(E(a))\leq \{ S_{\text{gen}}(W(a)),\ S_{\text{gen}}(a) \}$ are satisfied automatically. It leads to the following two constraints
\be
\begin{split}
&S_{\text{bulk}}(W(a))-S_{\text{bulk}}(W(a)\cup I(a)) \geq N^{i}_{1'}+N^{i}_{2'}=\frac{|\p I(a)|}{4G_N},\\
&S_{\text{bulk}}(a)-S_{\text{bulk}}(W(a)\cup I(a)) \geq N^{i}_{3}-N^{i}_{2}+N^{i}_{1'} \geq N^{i}_{3}-N^{i}_{2}-N^{i}_{2'} \geq \frac{|\p W(a)|+|\p I(a)|-|\p a|}{4G_N}.
\end{split}
\ee
Then by assuming the Araki-Lieb inequality of the bulk entanglement entropy, hence
\be
\begin{split}
&S_{\text{bulk}}(W(a)\cup I(a)) \geq S_{\text{bulk}}(W(a))- S_{\text{bulk}}(I(a)),\\
&S_{\text{bulk}}(a\cup (W(a)\setminus a) \cup I(a)) \geq S_{\text{bulk}}(a)- S_{\text{bulk}}((W(a)\setminus a) \cup I(a)),
\end{split}
\ee
we can find the lower bounds on the bulk entanglement entropy for region $I(a)$ and region $E(a)\setminus a=(W(a)\setminus a) \cup I(a)$ (with orientated boundary $\p (E(a)\setminus a)= \p W(a)+\p I(a)- \p a$) respectively as
\be \label{constraint2}
\begin{split}
&S_{\text{bulk}}(I(a)) \geq  \frac{|\p I(a)|}{4G_N},\\
&S_{\text{bulk}}(E(a)\setminus a) \geq  \frac{|\p W(a)|+|\p I(a)|-|\p a|}{4G_N}.
\end{split}
\ee

The constraints (\ref{constraint1}) for no-island phase and (\ref{constraint2}) for island phase are convincing, as we only utilized definitions of the GEW such that $S_{\text{gen}}(E(a))\leq \{ S_{\text{gen}}(W(a)),\ S_{\text{gen}}(a) \}$, and the Araki-Lieb inequality for the bulk quantum state of matter fields. They put constraints on the existence of a nontrivial GEW in general gravitational spacetimes, particularly the existence of an entanglement island, which may also be interpreted as the physical conditions for inducing
nontrivial quantum phase transitions in GEW prescription.

One may recall the studies on the entropy bounds, from the Bekenstein Bound \cite{Bekenstein:1980jp},  to the spherical entropy bound \cite{Susskind:1994vu}, the spacelike entropy bound \cite{Bousso:2002ju} and the covariant entropy bound \cite{Bousso:1999xy,Bousso:1999cb}, which put upper bounds on the bulk entanglement entropy of matter fields in certain regions. The first three entropy bounds suffered from several problems as stated in \cite{Bousso:2002ju}, while the covariant entropy bound has been proven to hold in a wide variety of situations \cite{Flanagan:1999jp,Wall:2010cj,Wall:2011hj,Bousso:2014sda}. In this paper, we are interested in the conjecture of spacelike entropy bound, which states that the entanglement entropy of matter fields contained in
any codimension-one spatial region $V$ will not exceed the boundary area of region $V$ in static Cauchy slices, that is
\be
S_{\text{bulk}}(V) \leq  \frac{|\p V|}{4G_N}.
\ee
In some special situations, the spacelike entropy bound may be valid for region $I(a)$, thus it forces that
\be
S_{\text{bulk}}(I(a))= \frac{|\p I(a)|}{4G_N},
\ee
by combining with constraint (\ref{constraint2}). However, we stress that this entropy bound can be violated even for isolated, spherical, weakly gravitating matter systems \cite{Bousso:2002ju}. As for the covariant entropy bound, it focuses on the entropy on the light-sheet $L(\p V)$ generated by null geodesics originating from codimension-two spacelike surface $\p V$, rather than the codimension-one spacelike region $V$, hence it may not be directly related to our results. However, under certain conditions, the covariant
bound indeed implies a spacelike bound by the spacelike projection theorem \cite{Bousso:1999xy,Bousso:1999cb,Bousso:2002ju}.

%%%%%%%%%%%%%%%%%%%%%%%%%%%%%%%%%%%%%%%%%%%%%%%%%%%%%%%%%%%%%%%%%%%%%%%%%%%%%%
\section{Conclusion and discussion}
\label{Sec:Discussion}
%%%%%%%%%%%%%%%%%%%%%%%%%%%%%%%%%%%%%%%%%%%%%%%%%%%%%%%%%%%%%%%%%%%%%%%%%%%%%%
In the paper, we showed that the GEW prescription implies the principle of the holography of information in topologically trivial Cauchy surfaces with a pure total state of matter fields, which indicates the holographic nature of the boundaries of these Cauchy slices. It prompted us to assume that the bit threads emerge from the boundaries of these Cauchy slices, like in AdS/CFT correspondence. Then we proposed the quantum bit thread formulation that is dual to the entropy formula from the GEW proposal by using the tools from convex optimization. In this way, we succeeded in extending the bit thread description to a static Cauchy slice in more general gravitational spacetimes, not limited to the AdS spacetime. By using the properties of flows, we proved the basic properties of the entropy for bulk gravitational regions, such as the monotonicity, subadditivity, Araki-Lieb inequality and strong subadditivity. We did not expect that the MMI property is satisfied for bulk regions in general gravitational spacetimes, but the MMI property may hold in some special gravitational spacetimes (such as AdS spacetime), then it may be possible to show the MMI property in these special gravitational spacetimes by introducing the so-called multiflow configurations \cite{Cui:2018dyq,Hubeny:2018bri,Agon:2021tia}. It would also be interesting to investigate whether there are certain conditions that lead to the existence of the MMI property from the GEW perspective. Furthermore, we found that the bulk entanglement entropy of matter fields in the region $E(a)\setminus a$ must be lower bounded by the area of the orientated boundary, as required by the existence of a nontrivial GEW on a static Cauchy slice in general gravitational spacetimes. In particular, the bulk entanglement entropy of matter fields on an entanglement island should be lower bounded by the boundary area of the entanglement island.

Note that we only investigated the bit thread formulation in static scenarios in the present paper. The generalizations to the Lorentzian and covariant settings like in \cite{Headrick:2017ucz, Headrick:2022nbe} would be worth studying, although a general time-dependent extension of GEW proposal \cite{Bousso:2022hlz,Bousso:2023sya} is still under study. Furthermore, though the GEW proposal provides a potential pattern of holographic encoding in general spacetimes, the fine-grained description of the entanglement structures still needs further studies. Since the bit thread description may help reveal more detailed structures of the entanglement entropy by connecting it with the information-theoretic contents as did in the AdS/CFT correspondence. For example, when the bulk quantum entanglement can be neglected, the entropy formula (\ref{genelized-QES}) becomes an RT-like formula. Meanwhile, the quantum bit thread description will reduce to the classical one that only contains the homogeneous threads, thus it would be easier to deal with. It may be feasible to introduce the corresponding bit thread description for other information-theoretic concepts (such as the entanglement of purification, partial entanglement entropy and multipartite entanglement) into general spacetimes.

Moreover, although we only considered a topologically trivial manifold $M$ and the total state of matter fields in $M$ is a pure state, we expect our bit thread description could be suitably extended to more general scenarios where $M$ has a non-trivial topology and the total state of matter fields in $M$ can be a mixed state. Note that there may exist a ``hole'' inside $M$, whose interior is not accessible for an outside observer on $M$. It allows the existence of an extra manifold $M'$ with another boundary $\p M'$ behind the ``hole'', where $M$ and $M'$ are smoothly joined together along the boundary of the ``hole'', such as a wormhole geometry. In fact, the topology of manifolds may be even more complicated. In addition, it is also possible for the total state of matter fields in $M$ to be a mixed state, as the matter fields in manifold $M$ may entangle with the matter fields in another manifold $M''$ with boundary $\p M''$, even though manifold $M''$ may not connect with manifold $M$ through any classical geometry. In short,
it means that some threads are allowed to connect region $a$ to these extra boundaries in homogeneous and inhomogeneous ways, thus we need also take this extra part of threads into consideration when maximizing the flux of any flow getting into region $a$. According to the entropy formula (\ref{genelized-QES}), we should minimize the generalized entropy among all homology regions of the region $a$, including the geometry and the entanglement entropy of matter fields on these extra manifolds. As long as we consider the whole manifold $M_{\text{total}}$ with its boundary $\p M_{\text{total}}$, the state of the matter fields in $M_{\text{total}}$ would be pure. By replacing $M$ with $M_{\text{total}}$ (hence $a^{c}\equiv M_{\text{total}}\setminus a$) and then maximizing the number of total threads from boundary $\p M_{\text{total}}$ to region $a$ in above formulation (\ref{quantum-maxflow-1}), our bit thread description would be applicable.

\section*{Acknowledgements}

This work is supported by the National Natural Science Foundation of China under Grant No. 11675272.

\appendix

\bibliographystyle{jhep}
\bibliography{refs}

\end{document}